\documentclass[%
 reprint,
superscriptaddress,
 amsmath,amssymb,
 aps,
]{revtex4-2}

\usepackage{float}
\usepackage{placeins}
\usepackage{graphicx}
\usepackage{dcolumn}
\usepackage{bm}
\usepackage{url} 
\graphicspath{{./UsedFigures/}}
\usepackage{comment}

\begin{document}

\title{Interpretable model-free inference of parametric variation across time-series data through large-scale feature extraction}


\author{Ben D. Fulcher$^\ast$}
\affiliation{School of Physics, The University of Sydney, Camperdown, 2006, NSW, Australia.}
\thanks{These authors contributed equally.}

\author{Carl H. Lubba$^\ast$}
\affiliation{Department of Bioengineering and Centre for Neurotechnology, Imperial College London, South Kensington, London SW7 2AZ, UK.}
\thanks{These authors contributed equally.}

\author{Giorgio F. Gilestro}
\affiliation{Department of Life Sciences, Imperial College London, South Kensington, London SW7 2AZ, UK.}

\author{Simon R. Schultz}
\affiliation{Department of Bioengineering and Centre for Neurotechnology, Imperial College London, South Kensington, London SW7 2AZ, UK.}

\author{Nick S. Jones}

\affiliation{Department of Mathematics and Centre for the Mathematics of Precision Healthcare, Imperial College London, South Kensington, London SW7 2AZ, UK.}

\date{\today}

\begin{abstract}
Here we address the problem of estimating the dimensionality and nature of parametric variation in an unknown generative process directly from time-series data, without specifying or fitting a model.
In particular we suppose that inter-instance variation in collections of time series is caused by parametric variation in the generating model.
We hypothesise that, given a sufficiently large library of time-series features, low-dimensional parametric variation will manifest as low-dimensional structure in feature space, enabling interpretable estimators of the underlying degrees of freedom to be constructed.
We test our hypothesis using a library of over 7000 diverse and interpretable time-series statistics and thirteen simulated systems with known parametric variation, spanning linear stochastic processes, nonlinear oscillators, and chaotic dynamics.
Our unsupervised, data-driven approach often reconstructs the underlying parametric variation across this extensive range of simulated dynamical systems while also yielding interpretable estimators for each underlying dimension.
Applied to the movement dynamics of 1143 fruit flies, we use this method to extract biologically meaningful components corresponding to sex and circadian rhythmicity.
Our results pave the way for much-needed data-driven methods to bridge the gap between interpretable theoretical understanding of dynamics and the large and complex datasets that characterize modern scientific problems.
\end{abstract}


\maketitle


\section{Introduction}

Many modern scientific challenges are characterized by an unprecedented scale of recorded time-series data and the corresponding development of new statistical tools to analyze them, with high-profile public datasets of real-world signals underlying progress in fields from biomedicine to astrophysics.
The hope of scientists aiming to distill interpretable understanding from large and complex datasets is that they are surprisingly `low-dimensional': they can be reduced to a smaller number of core, interpretable dimensions that capture much of the important variation across a dataset \cite{Hastie2009:ElementsStatisticalLearning}.

There are many senses in which a time-series dataset can be `low-dimensional'.
For a single multivariate time series, low-dimensional structure typically refers to reduced temporal components that are highly explanatory of the structure of the full dataset (e.g., using dimension-reduction methods like Principal Components Analysis (PCA) to capture variance explained, or other time-series-specific dimension-reduction methods to capture temporal structure~\cite{Owens2025:TimeseriesDimensionReduction}).
And in nonlinear time-series analysis, low-dimensional structure typically refers to the existence of a low-dimensional space, constructed through time-delay embedding~\cite{Takens81}, in which the underlying dynamical rules play out~\cite{Kantz04}.
Finding and analyzing explanatory low-dimensional temporal structures underlying a high-dimensional time series is a crucial component to modeling and understanding low-dimensional structure in a complex time-varying system.

Instead of inferring different types of low-dimensional structure from a single time series, here we consider the problem of inferring low-dimensional variation \textit{across a collection} of univariate time series.
In particular, we seek low-dimensional variation of dynamical properties across the set of time-series instances.
Such a method could help organize large databases of time series along meaningful dimensions.
For example, in a clinical setting, we would aim to organize a large set of physiological signals collected across a population along a reduced number of interpretable dimensions that best capture the inter-individual variation in the physiological dynamics.
And in simulated settings, we might use such a method to reconstruct the underlying free parameters of the model used to construct a time-series dataset, in the typical case where the variation of parameters drives differences in dynamical properties across time-series instances.
Furthermore, considering windows of a non-stationary process as a set of time-series instances (and ignoring their time ordering), the problem of tracking non-stationary variation from time-series data~\cite{Owens2024:ParameterInferenceNonstationarya} can also be adapted to this problem class.
This ability to represent a large dataset of complex dynamics succinctly, in terms of a reduced set of dynamical properties that capture variation across the dataset, is thus powerful and widely applicable for studying dynamical data.

To our knowledge, the closest prior work in this direction has been that of inferring the variation of some underlying parameter across time of some non-stationary process.
It has been shown that parameter variation in a non-stationary system can be tracked in the space of well-chosen time-series features that are sensitive to the dynamical changes produced by the underlying parameter~\cite{Guttler2001:ReconstructionParameterSpaces}.
But the success of this approach relies on the manual selection of time-series features that are sensitive to the underlying parameters, which is not straightforward without knowledge of the underlying system.

In this work we explore the hypothesis that a sufficiently comprehensive and diverse set of time-series features can capture arbitrary parametric variation, which could then be distilled via dimension reduction methods to infer the relevant dimensions of parameter variation underlying a time-series dataset, thereby bypassing the need for manual feature selection (as in \citet{Guttler2001:ReconstructionParameterSpaces}).
The relatively recent development of comprehensive sets of time-series features, like the \textit{hctsa} feature set containing over $7000$ time-series features~\cite{Fulcher2013:HighlyComparativeTimeseries, Fulcher2017:HctsaComputationalFramework}, allow us to operationalize this hypothesis.
In particular, \textit{hctsa} contains features that are sensitive to a wide range of statistical properties of time series, including properties of the distribution, linear and nonlinear correlations, spectra, information-theoretic properties, nonlinearity and chaos, self-affinity, stationarity, and motifs.
And, in the case that the data-generating process involves highly constrained parametric degrees of freedom (and where each parameter shapes a specific time-series property), we hypothesize that the recovered dimensions will capture the space of free parameters of the generative model.
We introduce a simple algorithm for inferring low-dimensional variation of dynamical properties across a time-series dataset by first projecting it to a high-dimensional feature space and then applying dimension reduction.
Using simulations of a diverse range of thirteen generative systems---from linear stochastic systems through to nonlinear deterministic chaotic systems, in each case generating a time-series dataset resulting from different degrees of parametric freedom (i.e., allowing a different number of parameters to vary freely)---we aim to test the hypothesis that low-dimensional structure of a time-series dataset represented in a space of extracted features corresponds to low-dimensional variation in the degrees of freedom of the generative process underlying the dataset.
We show that \emph{our library of features is sufficiently rich} to allow frequent recovery of the degrees of freedom in the generative model with high accuracy, while also providing interpretable statistical estimates of the underlying free parameters: making model-free exploratory analysis a credible route to characterizing generative structure.
Applying the method to a \textit{Drosophila} movement phenotyping dataset, we demonstrate its ability to extract informative dimensions of variation across an empirical dataset, with two key dimensions of inferred dynamical variation corresponding to: (i) sex differences; and (ii) time-of-day variation across the fly population.

\section{Methods}

In this section, we first introduce theory underlying the problem of inferring key dimensions over which statistical properties vary across a collection of univariate time series (in Sec.~\ref{sec:theory}), before explaining details of the methods for constructing the feature space and inferring low-dimensional structure in it (in Sec.~\ref{sec:feature_extraction_dim_red}), and then describing the set of model systems on which we evaluate our method (in Sec.~\ref{sec:simulated_datasets}).
Note that our library of datasets and \textit{hctsa} feature-extraction results are available at \url{https://figshare.com/s/b9ccfb28e6ac197fd31b} and analysis code to generate our results is available at \url{https://github.com/DynamicsAndNeuralSystems/TimeSeriesLowDimVariation_figures}.


\subsection{Theory}
\label{sec:theory}

We define a univariate time series, $x_t$ for $t = 1, 2, \dots, T$, as a uniformly-sampled, time-ordered sequence of $T$ measurements.
Henceforth we refer to this vector object as $\mathbf{x}$.
A time-series dataset, $\mathcal{X}$, is defined here as a collection of $N$ such univariate time series:
\begin{equation}
    \mathcal{X} = \{\mathbf{x}^{(1)}, \mathbf{x}^{(2)} ..., \mathbf{x}^{(N)}\}\,,
\end{equation}
where each time series, $\mathbf{x}^{(i)} \in \mathbb{R}^{T_i}$, is an ordered sequence of $T_i$ real-valued measurements in general (although we consider just the case of a common length $T$ here).

We consider each time series $\mathbf{x}$ as a realization from a common process $X_t$, $t = 1, \dots, T$, a sequence of random variables that defines a probability distribution over time-series instances $p(\mathbf{x})$.
We suppose that the generative process corresponds generally to some (unknown) time-series model $\mathcal{M}(\Theta)$ with behavior controlled by a set of $m$ free parameters, $\Theta = \{\theta_1, ..., \theta_m\}$.
Specifying the values of all parameters constrains the joint distribution over time series as $p(\mathbf{x} | \Theta)$ (and corresponds to a delta function centered on a specific sequence in the case of a deterministic model $\mathcal{M}$).
In general, the dynamical rules encoded in $\mathcal{M}$, and the joint distribution over parameters $p(\Theta)$ determines a distribution over time series, the process $[X_1,\dots,X_T]$, from which the time series $\mathbf{x}$ are sampled repeatedly to produce a time-series dataset $\mathcal{X}$.
In the case that the effect of each parameter on the dynamics can be effectively reduced to a single summary statistic, then the variation of dynamical properties observed across time series in $\mathcal{X}$ provides understanding of the free parameters $\Theta$ of the (unknown) data-generating process $\mathcal{M}$ (which together determine $p(\mathbf{x})$).


The aim of this work is to compute explanatory dimensions of variation of dynamical properties across a time-series dataset $\mathcal{X}$, as the basis of estimating the underlying space of parametric freedoms, $\Theta$, in the data-generating process.
Under the assumption that each relevant $\theta_i$ has a distinct effect on the statistical properties of the dynamics, $\mathbf{x}$, our proposed inference procedure involves first mapping each time series to a set of $P$ extracted real-valued features, $f_j$, for $j = 1, 2, \dots, P$, that we assume to be sufficiently comprehensive to contain features sensitive to the variation in time-series properties controlled by each free parameter $\theta_i$.
By representing all time series in the dataset, $\mathcal{X}$, as a set of $P$ features, the dataset can be rewritten as a set of observations in the $P$-dimensional feature space, i.e., as an $N \times P$ time series $\times$ feature matrix, $F$.


Our central hypothesis is that low-dimensional parametric variation $\Theta$ in the underlying model $\mathcal{M}$ will yield low-dimensional structure in the feature space $F$ \emph{if the feature space is sufficiently comprehensive}.
We will outline our argument for this claim by first considering the simple case: that variation in a single parameter $\Theta = \{ \theta \}$ is responsible for the variation in statistical properties of time series $\mathbf{x}^{(i)}$ in $\mathcal{X}$.
Then variation in all features $f_i$ can be attributed to either:
(i) stochasticity, in either the algorithm underlying the feature computation, or in the finite-sample time series, $\mathbf{x}^{(i)}$; or
(ii) a sensitivity of the statistical properties of the time series to variation in the underlying free parameter, $\theta$.
An independent stochastic source of feature variation across the dataset (e.g., a stochastic component of an algorithm) does not result in statistical dependence between features (and hence does not contribute to low-dimensional structure).
But if two or more features are sensitive to the variation in $\theta$, they will be distinguished through their covariation, leading to low-dimensional structure in the feature space that corresponds to the variation of $\theta$.
Extending this argument to the case of multiple free parameters is complicated in cases in which:
(i) groups of features in the feature set are correlated through their algorithmic construction, rather than due to structure in the dataset (i.e., large groups of features that measure common time-series properties bias low-dimensional components to capturing variance explained by these over-represented properties in the make-up of the feature set); and
(ii) the parameters affect statistical properties of time series in non-independent ways (leading to a blurring between low-dimensional components of dynamical variation in the dataset, which we estimate, and underlying parameter variation).
Even though both of these conditions are expected to apply in general and complicate the accuracy of our approach, here we test the hypothesis that for many different types of systems, the dominant low-dimensional structure in the feature space can nevertheless be used to reconstruct underlying parametric variation.

\begin{figure*}[ht]
\includegraphics[width=.95\textwidth]{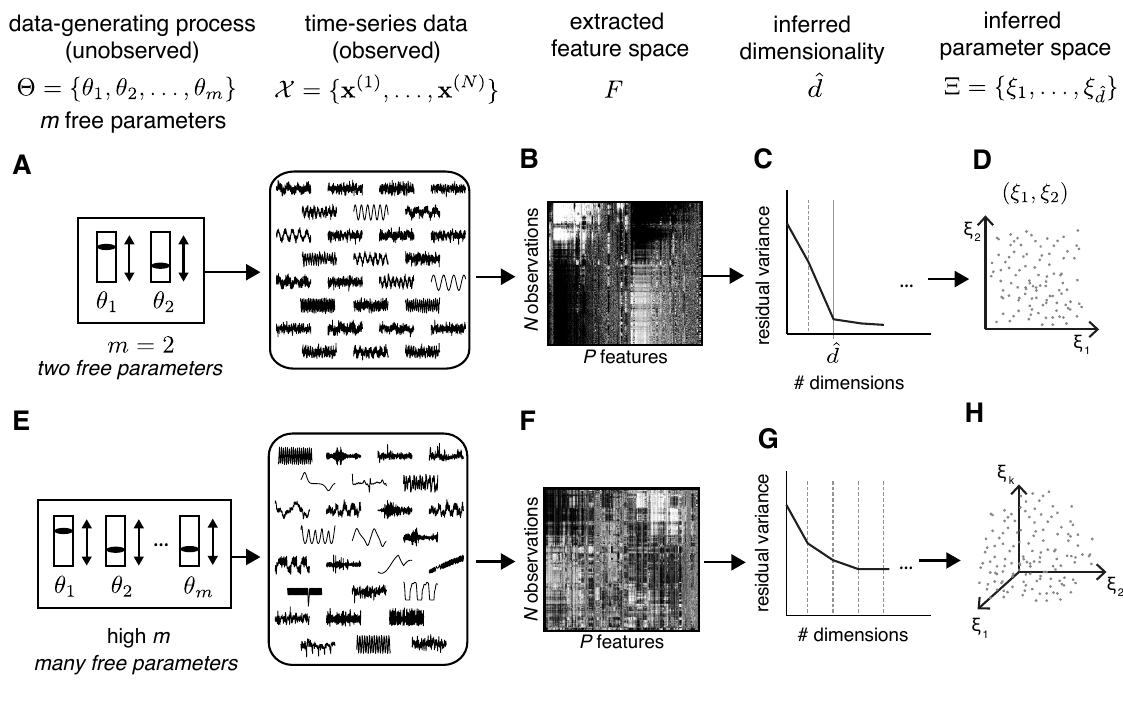}
\caption{
\textbf{Low-dimensional parametric variation of the generative process underlying a time-series dataset can be inferred by mapping each time series to an expressive space of time-series features and analyzing low-dimensional structure in the extracted feature space.}
We consider two cases: \textbf{A}--\textbf{D} a low-dimensional time-series dataset (generated from a noisy sinusoid model with two free parameters), and \textbf{E}--\textbf{H} a high-dimensional time-series dataset generated from a process with a very large number of free parameters.
\textbf{A},\textbf{E}: We first depict the data-generating process for each case, in which statistical variation in the set of time series results from variation from some number of free parameters in the underlying process, and is thus relatively constrained for the two-dimensional case (\textbf{A}) relative to the high-dimensional case (\textbf{E}).
\textbf{B},\textbf{F}: A comprehensive set of $P$ statistical features is extracted from each time series in a dataset, yielding a $N \times P$ feature matrix $F$, which is visually more low-dimensional in the case of the dataset simulated from the constrained model (\textbf{B}).
\textbf{C},\textbf{G}: Applying dimension-reduction to the feature space yields an estimate of its low-dimensional structure, which saturates at $\hat{d} = 2$ for the two-dimensional dataset (\textbf{C}), recapitulating the underlying number of free parameters in the generative process for $\mathcal{X}$.
\textbf{D},\textbf{H}: The estimated low-dimensional components $\Xi$ provide a data-driven embedding for the parameter space $\Theta$.
}
\label{fig:schematic}
\end{figure*}

The problem, and our approach to tackling it, is depicted schematically in Fig.~\ref{fig:schematic}.
As shown in Fig.~\ref{fig:schematic}A, for the purposes of demonstration we consider a noisy sinusoid model $\mathcal{M}$ with $m = 2$ free parameters $\Theta = \{\theta_1, \theta_2\}$ that control variation in the sinusoid frequency ($\theta_1$, say) and noise amplitude ($\theta_2$, say).
Each independent sample from these two parameters defines a time series, and a collection of time series $\mathcal{X}$ is generated through repeated sampling.
The variation in extracted time-series features across the dataset, visible through the covariation of large numbers of features in the feature matrix $F$ (in Fig.~\ref{fig:schematic}B) contains an imprint of the reduced parametric degrees of freedom in the generative model.
Specifically, features in $F$ either vary with: (i) $\theta_1$; (ii) $\theta_2$; (iii) both $\theta_1$ and $\theta_2$; or (iv) neither $\theta_1$ nor $\theta_2$, with inter-dependencies between features induced by $\theta_1$ and $\theta_2$ resulting in (generally nonlinear) low-dimensional structure in $F$.
Applying a nonlinear dimension-reduction algorithm to $F$, shown in Fig.~\ref{fig:schematic}C, reveals a saturation in residual variance explained in $F$ at $\hat{d} = 2$, with the resulting two estimated dimensions, $\xi_1$ and $\xi_2$, yielding candidate estimates for the underlying free parameters $\theta_1$ and $\theta_2$.
In this way, we obtain a data-driven reconstruction of the parameter space of the (in general unknown) data-generating model, $\Theta = (\theta_1, \theta_2)$ as $\Xi = (\xi_1, \xi_2)$, depicted in Fig.~\ref{fig:schematic}D.

This result for a dataset produced from low-dimensional parameter variation, can be contrasted with a generative model with many more free parameters, shown in Fig.~\ref{fig:schematic}E, which yields time series with a far wider range of dynamical properties than the highly constrained noisy sinusoid model.
In this case, each feature $f_i$ varies in a complicated way with the variation in many more underlying parameters, $\{\theta_1, \dots, \theta_m\}$, yielding a higher-dimensional space of estimated statistical properties, $F$ (Fig.~\ref{fig:schematic}F), that can again be quantified using a dimension-reduction algorithm (Fig.~\ref{fig:schematic}G).
A low-dimensional reconstruction may capture some dominant sources of variation (Fig.~\ref{fig:schematic}H) but cannot fully reconstruct the complex high-dimensional parametric variation.

\subsection{Feature extraction and dimension reduction}
\label{sec:feature_extraction_dim_red}

The ability to infer the variation in time-series statistics resulting from variation of a set of free parameters $\Theta$ in an unknown model $\mathcal{M}$ requires the feature set $\{f_i\}$ to be sufficiently comprehensive to contain statistical estimators that are sufficiently sensitive to the statistical variation.
In their work on using time-series summary statistics to track parameter variation, \citet{Guttler2001:ReconstructionParameterSpaces} acknowledged the difficulty of being able to track a given parameter from time-series data in general, and focused on the case in which such estimators need to be selected manually (based on knowledge of the underlying system).
Here we take a data-driven route to identifying relevant time-series summary statistics by drawing on a comprehensive set of $>7000$ time-series features in the \textit{hctsa} feature library~\cite{Fulcher2017:HctsaComputationalFramework}.
The \textit{hctsa} feature set quantifies a wide range of time-series properties, including properties of the distribution of values, linear and nonlinear autocorrelation structure, stationarity, summaries of basis-function decompositions (including Fourier and wavelet transforms), fits to various time-series models (e.g., autoregressive, state space, Gaussian Process, and hidden Markov models), measures from nonlinear time-series analysis (e.g., correlation-dimension estimates, nonlinear prediction errors, fractal scaling properties, etc.), and information theoretic complexity and entropy measures.
Here we used version 0.97 of \textit{hctsa} \cite{Fulcher2017:HctsaComputationalFramework}.

After computing a feature matrix $F$ with which to represent the dynamical properties of time-series dataset $\mathcal{X}$, we normalized each column to a common scale using a robust sigmoid transform \cite{Fulcher2013:HighlyComparativeTimeseries}, to avoid dependence of the results on the arbitrary range over which different features vary (while also minimizing the impact of any large outliers).
Features that were constant across the dataset or were inappropriate for the data (such as features that attempt to fit a positive-only distribution to non-positive data) were removed.
These normalization and filtering procedures yielded a reduced, normalized time-series feature matrix $\tilde{F}$ for each dataset.

To quantify low-dimensional structure in $\tilde{F}$, here we used Isomap, which is sensitive to nonlinear structure (and used the default number of seven neighbors to create a neighborhood graph)~\cite{Tenenbaum2000:GlobalGeometricFramework}.
We compared the behavior of Isomap to a linear alternative, PCA, in Sec.~\ref{sec:robustnsss}, finding broadly similar (but overall inferior) reconstruction performance.
We also expect broadly similar results with other dimension-reduction methods, but note that our goal here is not to identify the optimal approach for manifold learning but rather to illustrate the general efficacy of dimension-reduction approaches in this context.
The results of dimension reduction yields a representation of the time-series dataset in a lower, $\hat{d}$-dimensional space $\Xi = (\xi_1, \xi_2, \dots, \xi_{\hat{d}})$ that aims to capture the dominant dimensions of variation in the high-dimensional feature space $\tilde{F}$.
In order to estimate the relevant dimensionality $\hat{d}$, we computed a measure of residual variance $\sigma^2_\mathrm{resid}$ at each dimensionality $d$.
Residual variance was calculated as
\begin{equation}
    \sigma^2_\mathrm{resid}(d) = 1 - [r(D_F, D_\Xi)]^2\,,
\end{equation}
where the Pearson correlation $r$ is computed over the set of all unique pairwise distances $D_F$ (in the normalized feature space $\tilde{F}$) and $D_\Xi$ (in the low-dimensional space $\Xi$).
Then $\hat{d}$ was estimated on the basis of the saturation of $\sigma^2_\mathrm{resid}(d)$, computed here using a heuristic that involved first computing $d_\mathrm{sat}$ as the minimal $d$ for which $\sigma^2_\mathrm{resid}(d + 1) - \sigma^2_\mathrm{resid}(d) < \epsilon_1$, for a small threshold parameter, $\epsilon_1$.
Then $\hat{d}$ was estimated as the minimal $d$ for which $\sigma^2_\mathrm{resid}(d) < \sigma^2_\mathrm{resid}(d_\mathrm{sat}) - \epsilon_2$, for a second small threshold parameter, $\epsilon_2$.
The two small threshold parameters used here, $\epsilon_1 = 0.0025$ and $\epsilon_2 = 0.05$, were set manually from the decay characteristics of $\sigma^2_\mathrm{resid}$ that we observed across the simulated systems studied here.


To measure how well the data-inferred dimensions $\Xi$ capture the ground-truth parametric variation $\Theta$ of a given simulated system, we attempted to reconstruct each of the original parameters, $\theta_1, \dots, \theta_m$, as a linear combination of the reduced dimensions, as $\hat{\theta}_i = \sum_{j = 1}^d a_j \xi_j$.
For each model parameter $\theta_i$, we quantified the reconstruction accuracy as the unexplained variance $\chi = 1 - R^2$ between the variation of that parameter across the dataset (for Pearson correlation coefficient $R$) and this linear estimate of it $\hat{\theta}_i$ for a given dimensionality $d$.
Note that, in our measures of $\chi$ involve a linear mapping and compute a linear correlation between $\Theta$ to $\Xi$, such that the resulting $\chi$ estimates can be considered an underestimate (relative to nonlinear transformations/correlations) of the correspondence, that is nevertheless sufficient for our purposes.


\subsection{Simulated datasets}
\label{sec:simulated_datasets}

To evaluate our methodology, we simulated time-series datasets from thirteen different time-series models, $\mathcal{M}$, covering different classes of dynamics:
\textit{linear and nonlinear stochastic systems} (autoregressive process \cite{Chatfield04},
noisy sinusoid with linear trend and mean shift,
a noisy bimodal switching system,
a predator-prey system \cite{May1972},
population growth \cite{verhulst1845loi}),
a \textit{deterministic oscillator} (van der Pol oscillator \cite{VanderPol1926}),
\textit{deterministic low-dimensional chaotic maps} (stochastic sine map \cite{Freitas2009:FailureDistinguishingColored}, Logistic Map \cite{May1976}) and flows (Lorenz \cite{Lorenz1963} and R\"ossler attractors \cite{Rossler1976}),
a \textit{high-dimensional chaotic flow} (the nonlinear time-delay Mackey--Glass equation~\cite{Mackey1977:OscillationChaosPhysiological}),
and a model of \textit{self-affine} time series~\cite{malamud1999self}.
Each time-series dataset $\mathcal{X}$ from a given model $\mathcal{M}$ was generated by independently sampling from a probability distribution over that system's free parameters $\Theta$.
That is, each sample from $\Theta$ was used to generate a time series $x_t$, and repeated sampling from $\Theta$ yielded a time-series dataset $\mathcal{X}$.
Full details of the time-series models used and how they were simulated, are in Appendix~\ref{sec:all_systems}.

Here we aimed to investigate the ability of the method described above to capture low-dimensional parameter variation from simulated systems directly from a dataset $\mathcal{X}$ (i.e., without knowledge of the system).
We focus on datasets produced through random variation of one, two, or three free parameters, corresponding to $\Theta = \{\theta_1\}$, $\Theta = \{\theta_1, \theta_2\}$, or $\Theta = \{\theta_1, \theta_2, \theta_3\}$, respectively.
We used a uniform distribution with a specified range over which to independently sample free parameters, while fixing the values of other parameters (see Sec.~\ref{sec:all_systems} for ranges and fixed values used).
For models with multiple free parameters, we could generate multiple datasets, corresponding to variation of a single free parameter (datasets labeled $\mathcal{X}^{(1)}$), various combinations of two free parameters ($\mathcal{X}^{(2)}$), and from allowing all three parameters to vary independently ($\mathcal{X}^{(3)}$).
In total, we generated 93 time-series datasets containing over 99\,000 time series:
46 datasets of $\mathcal{X}^{(1)}$ (100 time series per dataset);
37 datasets of $\mathcal{X}^{(2)}$ (400 time series per dataset); and
10 datasets of $\mathcal{X}^{(3)}$ (8000 time series per dataset).
We set a minimum dataset size of 100 time series, as used for $\mathcal{X}^1$, and for $\mathcal{X}^2$ and $\mathcal{X}^3$ we used 20 time series per degree of freedom, resulting in simulations of 400 and 8000 time series for the two cases, respectively.

\begin{figure}[!htb]
\includegraphics[width=\columnwidth]{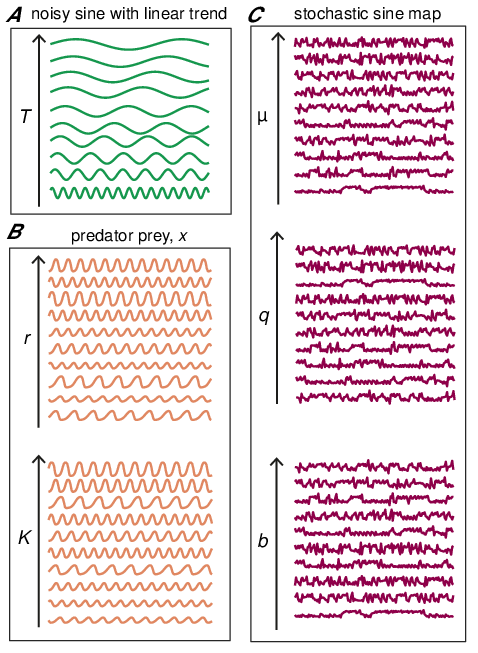}
\caption{
\textbf{An example time-series dataset generated from systems with one, two, or three free model parameters.}
For each dataset, we plot segments (the first 150 samples) of 10 randomly selected time series, ordered by one of the model parameters used to generate the dataset.
\textbf{A} An example dataset generated from variation of one free parameter is the `noisy trendy sinusoid' model, Eq.~\eqref{eqn:noisy_trendy_sine}, where the noise level, $\eta = 0$, and trend, $m = 0$, are fixed, and the variation in statistical properties across the dataset is underpinned by variation in the single free parameter, the sinusoidal period $T$.
There is a clear visual signature of the underlying one-dimensional parametric variation in $T$.
\textbf{B}
An example dataset generated from independent variation of two free parameters is the predator--prey model, for which both the growth rate $r$ and carrying capacity $K$ vary across the dataset.
\textbf{C}
An example dataset generated from independent variation of three free parameters is the stochastic sine map, for which the parameter values of $\mu$, $q$, and $b$ all vary across time series.
}
\label{fig:examplesMeasurementSpace}
\end{figure}

To give a visual intuition for time-series datasets generated from variation of one, two, or three free parameters ($\mathcal{X}^{(1)}$, $\mathcal{X}^{(2)}$, and $\mathcal{X}^{(3)}$), some example time-series segments from one dataset of each of these three types are plotted in Figs~\ref{fig:examplesMeasurementSpace}A--C, respectively.
The `noisy trendy sine model', Eq.~\eqref{eqn:noisy_trendy_sine}, has three parameters corresponding to the gradient of a linear trend, $m$, level of additive noise, $\eta$, and period of the sinusoidal signal, $T$.
Fixing both the noise level $\eta = 0$ and trend $m = 0$, and sampling from the free parameter, the period $T$, results in each time series being generated deterministically as $x_t = \sin(2\pi t/T)$.
The resulting dataset is generated from repeated random sampling of the free parameter $T$; 10 randomly selected time-series segments are shown in Fig.~\ref{fig:examplesMeasurementSpace}A, ordered by $T$.
In this case, the goal of the inference problem is to reconstruct the one-dimensional variation underlying the statistics of the dataset, and infer a statistical estimator for $T$, directly from the time-series data and without any knowledge of the generative process.

Relative to the visually clear one-dimensional variation in Fig.~\ref{fig:examplesMeasurementSpace}A, the variation is far more subtle in general (e.g., when the free parameter affects less obvious dynamical properties in a time series or when multiple parameters vary simultaneously).
For example, Fig.~\ref{fig:examplesMeasurementSpace}B shows example time series from a two-dimensional dataset, $\mathcal{X}^{(2)}$, generated by independently sampling the parameters $r$ (growth rate) and $K$ (carrying capacity) of a predator--prey model, Eq.~\eqref{eqn:predator_prey}~\cite{May1972, Hoppensteadt:2006}.
A sample of time series are ordered by both $r$ (upper), revealing a characteristic variation in waveform shape, and $K$ (lower), revealing a variation in signal amplitude.
We have selected this example for its visual clarity, but variation in the parameter of interest is more typically obscured by the independent variation of the other parameter in $\mathcal{X}^{(2)}$ datasets.

Finally we show example time series generated from the stochastic sine map model with three free parameters $\Theta = (\mu, q, b)$, in Fig.~\ref{fig:examplesMeasurementSpace}C.
Even when ordering time series by a ground-truth parameter, the independent variation of the other two free parameters obscures any clear visual effect of any single parameter on the statistics of the time series.
This is typical of the challenge in inferring the parametric variation underlying datasets generated by the variation of multiple free parameters, as in the $\mathcal{X}^{(3)}$ datasets.

\section{Results}

Using a range of simulated systems with controlled parameter variation, we aim to evaluate the ability of our simple data-driven method (based on estimating low-dimensional structure in a high-dimensional space of extracted time-series features) to infer the parametric variation in the underlying generative model across a wide range of simulated systems.

\subsection{A sufficiently comprehensive time-series feature library}

In order to yield corresponding feature--feature correlations in $F$ that are indicative of low-dimensional structure, the success of our approach require a comprehensive library of time-series features that contains multiple statistical estimators sensitive to each type of parametric variation in a given model.
We thus first aimed to assess whether the \textit{hctsa} feature set used here is sufficiently diverse to capture the variation in statistical properties of time series due to each underlying parameter in the time-series models analyzed.
For each parameter in each dataset we computed $\max(|\rho|)$, the maximum absolute Spearman correlation coefficient between the variation of the parameter with any feature in \textit{hctsa}.
A high $\max(|\rho|)$ value indicates the presence of at least one time-series feature in \textit{hctsa} that varies strongly and monotonically with the underlying parameter (despite confounding variation of at least one other parameter in the case of $\{\mathcal{X}^{(2)}\}$ and $\{\mathcal{X}^{(3)}\}$ datasets, cf. Fig.~\ref{fig:examplesMeasurementSpace}).

\begin{figure}[htb]
\includegraphics[width=\columnwidth]{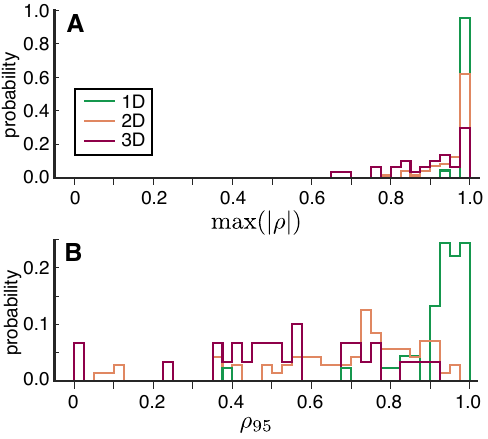}
\caption{
\textbf{
The \textit{hctsa} feature set is sufficiently comprehensive to capture the variation in time-series dynamics produced by parameters of a wide range of time-series models.
}
We plot the distribution of:
\textbf{A}: $\max(|\rho|)$, the maximum absolute correlation between the variation of a given parameter across a dataset, and that of each feature;
and
\textbf{B}: $\rho_{95}$, the 95th percentile of the distribution of correlations (a high $\rho_{95}$ value indicates the presence of many features sensitive to the parameter of interest).
Distributions are shown for:
46 one-parameter datasets, $\mathcal{X}^{(1)}$ (green),
74 parameters across 37 two-parameter datasets, $\mathcal{X}^{(2)}$ (orange),
and 30 parameters across 10 three-parameter datasets, $\mathcal{X}^{(3)}$ (purple).
}
\label{fig:feasibilityonly}
\end{figure}

Distributions of $\max(|\rho|)$ for one-, two-, and three-dimensional datasets are plotted in Fig.~\ref{fig:feasibilityonly}A.
When only a single parameter varies (one-dimensional parametric freedom), $\max(|\rho|) > 0.9$ for all parameters in all systems, indicating the presence of at least one statistical estimator in \textit{hctsa} that is able to monotonically capture the parameter variation.
As expected, when there is confounding variation in a different parameter (as in datasets generated from two or three free parameters, $\mathcal{X}^{(2)}$, $\mathcal{X}^{(3)}$), isolating the variation in any single parameter becomes more challenging, although the existence of an informative parameter (typically with $\max(|\rho|) > 0.8$) remains.
Each parameter of every system was well captured by at least one feature, at a correlation of at least 0.95 (across all $\mathcal{X}^1$ datasets), 0.79 (across all $\mathcal{X}^2$ datasets), and 0.66 (across all $\mathcal{X}^3$ datasets).
This result demonstrates the diversity contained in the \textit{hctsa} feature library in being able to track the wide range of parametric freedoms represented across our diverse set of time-series models.

Since the success of our unsupervised inference method requires multiple features sensitive to each potential source of underlying parameter variation across a dataset (in order to drive low-dimensional structure), we next tested for the existence of a set of features jointly sensitive to a given parameter variation in each case.
For this purpose we used $\rho_{95}$, the 95th percentile of the distribution of correlations $|\rho|$ (rather than the maximum, as above).
Distributions of $\rho_{95}$ are plotted separately for one-, two-, and three-dimensional datasets in Fig.~\ref{fig:feasibilityonly}B.
High $\rho_{95}$ values were observed in $\mathcal{X}^{(1)}$ datasets generated from variation of a single parameter, with a median $\rho_{95}$ of 0.94 across 46 such datasets.
As above, this demonstrates the feasibility of detecting these parametric constraints in an unsupervised way as low-dimensional structure in the feature space.
We note two poor-performing parameters with $\rho_{95} < 0.7$: the mean-shift parameter, $s$, in the `noisy shifty sine' model, Eq.~\eqref{eqn:noisy_shifty_sine} ($\rho_{95} = 0.38$), and the linear trend parameter, $m$, in the `noisy trendy sine' model, Eq.~\eqref{eqn:noisy_trendy_sine} ($\rho_{95} = 0.68$).
In both cases, \textit{hctsa} contains multiple features that correlate strongly with the relevant parameters (e.g., the mean estimator has a correlation $\rho = 1.00$ with the mean-shift, $s$, and a measure of linear gradient has a strong $\rho = 0.98$ with the trend gradient $m$), but such simple features are relatively rare in \textit{hctsa}, which is dominated by dynamical statistics of normalized time-series data.
Due to the make-up of \textit{hctsa}, such sources of dynamical variation, to which only a small fraction of the feature library is sensitive, are expected to be more challenging to detect as low-dimensional structure by our method.

The difficulty in tracking a given parameter increases substantially when there is additional independent (confounding) variation from other parameters: $\rho_{95}$ drops to 0.74 for $\mathcal{X}^{(2)}$ datasets in which there is confounding variation of another parameter (74 parameters across 37 datasets),
and drops to 0.53 when there is confounding variation of two other parameters (30 parameters across 10 $\mathcal{X}^{(3)}$ datasets).
For some parameters in some systems, we found high $\rho_{95}$ for $\mathcal{X}^{(1)}$ but a substantial drop in the $\mathcal{X}^{(2)}$ setting, due to the new parameter dominating the empirical dynamics relative to the parameter of interest.
An example is the noise-switch parameter $q$ in the stochastic sine map, Eq.~\eqref{eqn:stochastic_sine_map}, which injects random noise at a given time step with probability $q$.
This parameter is well estimated by individual features when only $q$ is freely varying ($\rho_{95} = 0.99$), but much less well ($\rho_{95} = 0.50$) when the sine map parameter $\mu$, which dominates the dynamics, varies independently.
Overall, we found strong support for \textit{hctsa} being a comprehensive interdisciplinary feature library for detecting diverse parametric variation across the broad range of dynamical models analyzed here.

\subsection{Case study: The van der Pol Oscillator}
\label{sec:van_der_pol}

\begin{figure*}[htb]
\includegraphics[width=\textwidth]{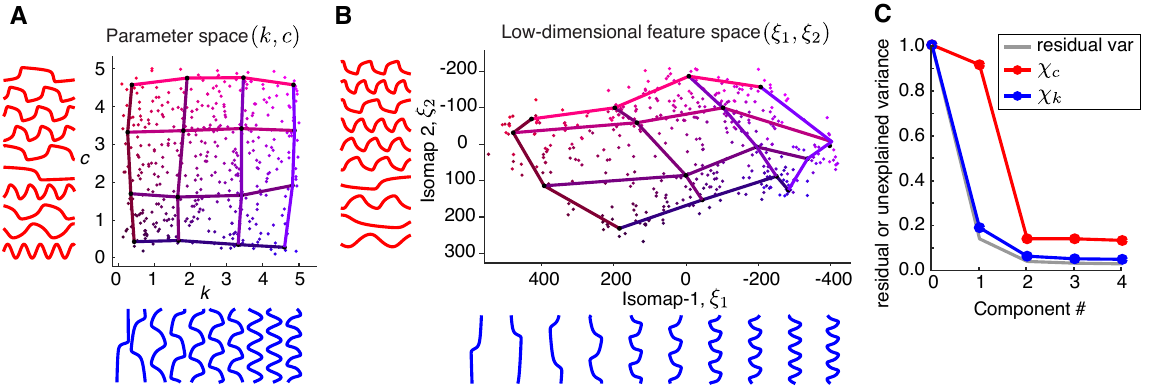}
\caption{
\textbf{Our method successfully infers low-dimensional parametric variation underlying a time-series dataset constructed by independently varying two parameters of the van der Pol oscillator.}
\textbf{A} The dataset contains 400 time series, each of which was generated from a point in the two-dimensional parameter space $(k,c)$, shown here.
Points are colored by their position in the space, from black $\rightarrow$ blue along $k$ and from black $\rightarrow$ magenta along $c$.
Some example time-series segments are plotted along the horizontal and vertical axes, ordered by increasing $k$ and $c$, respectively.
To aid visualization, annotated lines connect points in the space along which a given parameter (either $c$ or $k$) is approximately constant.
\textbf{B}
Our unsupervised reconstruction of the parametric variation, $(c,k)$, as $(\xi_1,\xi_2)$, is based on a low-dimensional projection of the dataset in a high-dimensional normalized feature space.
Points are colored according to their ground-truth parameter values, $(k,c)$, as in A, revealing a good reconstruction.
\textbf{C}
Unexplained variance is plotted as a function of the number of dimensions, $\hat{d}$ of $\Xi$ included in the reconstruction, as a gray line (saturating at $\hat{d} = 2$, the dimensionality of the underlying parameter space).
We also show the unexplained variance in reconstructing $c$ and $k$ from a given number of components, $\hat{d}$, as $\chi_c$ and $\chi_k$ (with $\xi_1$ mostly capturing variation in $k$ and $\xi_2$ mostly capturing variation in $c$).
}
\label{fig:casestudyVdP}
\end{figure*}

Before examining the behavior of our method across all systems, it is illustrative to walk through a demonstrative, well-behaved example of the two-parameter van der Pol oscillator.
The dynamics of the variable $x$ is governed by
\begin{equation}
\ddot x - c(1 - x^2) \dot x + kx = 0\,,
\end{equation}
with dynamical behavior modulated by two parameters $\Theta = (c,k)$.
We generated a $\mathcal{X}^{(2)}$ dataset of 400 time series by allowing both $c$ and $k$ to vary independently: sampling $c \sim U(0.1,5)$ and $k \sim U(0.1,5)$ from uniform distributions.
The sampled points in the $(c,k)$ parameter space are plotted in Fig.~\ref{fig:casestudyVdP}A with some example time-series segments ordered by increasing $c$ and $k$ plotted along each axis.
We applied our method to the set of 400 time series (and without any knowledge of the generative model or its parameters).

The resulting low-dimensional space $(\xi_1, \xi_2)$ captures the main sources of variation in dynamical properties of the time series across the dataset, as shown in Fig.~\ref{fig:casestudyVdP}B.
The plot shows every time series embedded as a point in the two-dimensional space, which clearly recapitulates the $(k,c)$ space of parametric variation of the underlying van der Pol model.
The residual variance, $\sigma^2$, as a function of the number of Isomap dimensions, $\hat{d}$, is plotted gray in Fig.~\ref{fig:casestudyVdP}C.
We see saturation at $\hat{d} = 2$, supporting the two-dimensional $(k,c)$ parameter variation used to construct the dataset.
Figure~\ref{fig:casestudyVdP}C also plots the unexplained variance, $\chi_k$ and $\chi_c$, in $k$ and $c$, respectively.
We see that the first component $\xi_1$ acts as a data-driven estimate for $k$ ($\chi_k$ drops at $\hat{d} = 1$) while the second component $\xi_2$ acts as a data-driven estimate of $c$ (since $\chi_c$ drops at $\hat{d} = 2$).
After two dimensions, both parameters are captured well ($\chi_k = 0.07$ and $\chi_c = 0.09$) with minimal improvement thereafter, consistent with the two-parameter $(k,c)$ space that underlies variation across the dataset.


An advantage of using time-series features derived from interpretable scientific theory is that it allows us to interpret the features that mostly strongly contribute to the extracted dimensions, $\xi_1$ and $\xi_2$.
As a simple demonstration of this, we inspected the list of features that exhibited the strongest correlation to $\xi_1$ and $\xi_2$.
Consistent with $k$ controlling oscillation frequency, $\xi_1$ is highly correlated to time-series features derived from the power spectrum (that provide good empirical estimates of the oscillation frequency).
And consistent with $c$ affecting the shape of the waveform, $\xi_2$ is highly correlated to features measuring smoothness/sharpness characteristics of the waveform (using momentum and wavelet-based features that act as strong individual estimates of $c$).


To summarize, applied to the van der Pol oscillator system, our data-driven method:
(i) produces a structured representation of the dataset in a low-dimensional embedding space that approximately reconstructs the parameter space sampled by the underlying generative model;
(ii) correctly estimates the two-dimensional parametric variation underlying variation across the dataset; and
(iii) highlights interpretable estimators of those two dimensions (features derived from a Fourier transform, sensitive to variation in oscillation frequency governed by $k$, and features quantifying waveform shape, governed by the model parameter $c$).
The strong performance of our method is underpinned by:
(i) the feature-based representation of the dataset (e.g., applying dimension reduction in the raw measurement space, the 400 observations $\times$ time point matrix, failed, with $\chi_c = 0.97$ and $\chi_k = 0.79$); and
(ii) the comprehensiveness of the \textit{hctsa} feature space (e.g., repeating using a smaller set of 22 time-series features~\cite{Lubba:2019} fails to capture the subtle variations in waveform shape, with $\chi_c > 0.9$).

\subsection{Low-dimensional projections}

\begin{figure*}[!htb]
\includegraphics[width=\textwidth]{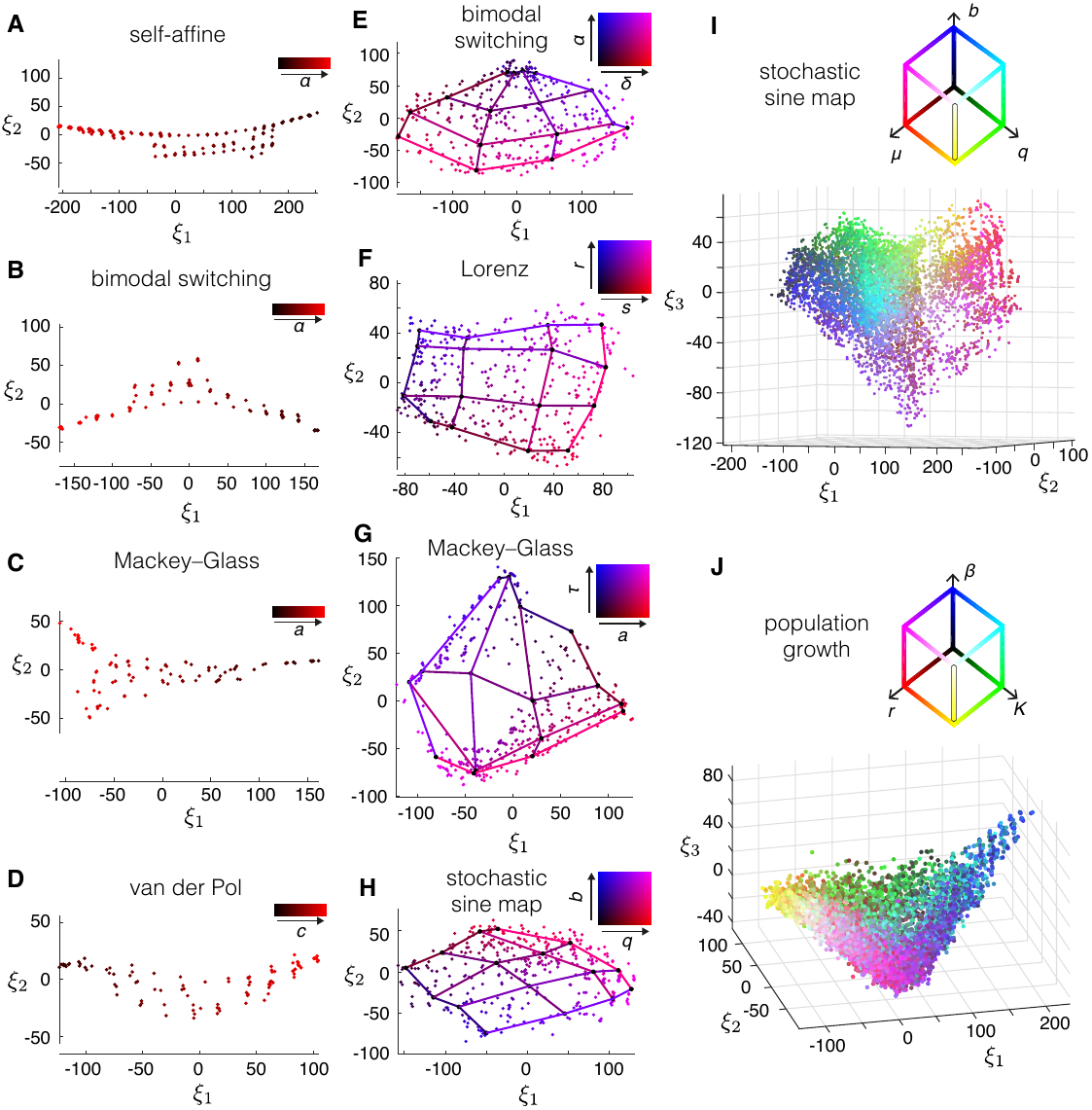}
\caption{
\textbf{Our data-driven method allows underlying parameter spaces of underlying generative models to be reconstructed from unlabeled time-series data.}
This figure shows selected examples of cases in which our extracted dimensions $\Xi$ successfully capture parameter variation underlying diverse systems with:
one free parameter, $\mathcal{X}^{(1)}$ (\textbf{A}--\textbf{D});
two free parameters, $\mathcal{X}^{(2)}$ (\textbf{E}--\textbf{H}); and
three free parameters, $\mathcal{X}^{(3)}$ (\textbf{I} and \textbf{J}).
Results are shown for the first two extracted dimensions $(\xi_1,\xi_2)$ for $\mathcal{X}^{(1)}$ and $\mathcal{X}^{(2)}$ datasets, and for the first three dimensions $(\xi_1,\xi_2,\xi_3)$ for the two $\mathcal{X}^{(3)}$ datasets.
Systems are as follows, specifying the system variable (where appropriate) and free parameter(s) in each case:
\textbf{A}: Self-affine \cite{malamud1999self};
\textbf{B}: Bimodal: $x$, $(\alpha)$ \cite{Fulcher2012};
\textbf{C}: Mackey--Glass $(a)$ \cite{Glass1988};
\textbf{D}: van der Pol: $x$, $(c)$ \cite{VanderPol1926};
\textbf{E}: Bimodal $(\delta, \alpha)$ \cite{Fulcher2012};
\textbf{F}: Lorenz: $y$, $(s,r)$ \cite{Lorenz1963};
\textbf{G}: Mackey--Glass $(a,\tau)$ \cite{Glass1988};
\textbf{H}: Stochastic sine map, $(q,b)$ \cite{Freitas2009:FailureDistinguishingColored};
\textbf{I}: Stochastic sine map $(\mu,q,b)$ \cite{Freitas2009:FailureDistinguishingColored}; and
\textbf{J}: Population growth $(r,K,\beta)$ \cite{Levins1969}.
Each plotted point represents a time series in an embedded space and is colored according to the underlying parameter values, according to an annotated color map for labeling parameter values in one-dimensional, two-dimensional, and three-dimensional parameter spaces.
For the two-dimensional parameter spaces inferred in panels \textbf{E}--\textbf{H}, grid lines are also annotated, connecting points in the space along which a given parameter is approximately constant.
}
\label{fig:goodcases}
\end{figure*}

Having characterized the performance of our method on an exemplar two-parameter system, the van der Pol oscillator, we next highlight examples from a wider range of systems on which our method performed well, shown in Fig.~\ref{fig:goodcases}.
As shown in Figs~\ref{fig:goodcases}A--D, for datasets generated through varying a single parameter, $\mathcal{X}^{(1)}$, our method yielded approximately one-dimensional structures in the extracted embedding space, $\Xi$.
Such variation could be inferred despite a diverse types of dynamical variation, generated by:
varying the powerlaw scaling exponent, $\alpha \sim U(-1,3)$, in a dataset of self-affine time series (Fig.~\ref{fig:goodcases}A);
varying the switch-rate, $\alpha \sim U(0,1)$, in a bimodal switching model, Eq.~\eqref{eqn:bimodal_switching} (Fig.~\ref{fig:goodcases}B);
varying the parameter $a \sim U(0.15,1.5)$ of the Mackey--Glass system, Eq.~\eqref{eqn:mackey_glass} \cite{Mackey77} (Fig.~\ref{fig:goodcases}C);
and varying the shape parameter, $c \sim U(0.1,5)$ of the van der Pol oscillator, Eq.~\ref{eqn:vanderPol} \cite{Sprott03} (Fig.~\ref{fig:goodcases}D).
In each case, $\xi_1$ provides an estimate of the underlying free underlying parameter $\theta_1$.

When two parameters vary independently, as in the $\mathcal{X}^{(2)}$ datasets, the underlying two-dimensional parameter space is approximately reconstructed in the extracted embedding (as per the van der Pol oscillator studied in Sec.~\ref{sec:van_der_pol} above).
Examples are plotted for datasets generated by:
varying the well separation $\delta \sim U(0,6)$ and switch probability $\alpha \sim U(0,1)$ in the bimodal switching model, Eq.~\eqref{eqn:bimodal_switching} (Fig.~\ref{fig:goodcases}E);
varying $s \sim U(8,30)$ and $r \sim U(35,60)$ of the Lorenz attractor, Eq.~\eqref{eqn:lorenz} (Fig.~\ref{fig:goodcases}F);
varying $a \sim (0.15, 1.5)$ and $\tau \sim \{10, 11, ..., 40\}$ in the Mackey--Glass system, Eq.~\eqref{eqn:mackey_glass} (Fig.~\ref{fig:goodcases}G);
and varying the noise occurrence probability $q\sim U(0,1)$ and amplitude $b\sim U(0,3)$ of the stochastic sine map, Eq.~\eqref{eqn:stochastic_sine_map} (Fig.~\ref{fig:goodcases}H).

As expected, the degree to which the underlying rectangular domain sampled in the parameter space is reproduced as a corresponding rectangular geometry in the low-dimensional feature space reconstruction differs across the systems.
The rectangular grid is well reconstructed for the Lorenz attractor and stochastic sine map (Figs~\ref{fig:goodcases}F,H), but with some level of distortion for the bimodal switching system (Fig.~\ref{fig:goodcases}E) and Mackey--Glass system (Fig.~\ref{fig:goodcases}G).
For example, at low values of $\delta$, the bimodal switching system contracts towards a point (Fig.~\ref{fig:goodcases}E), suggesting that the other free parameter $\alpha$ has a large effect on the dynamics when $\delta$ is large and a vanishingly small effect as $\delta$ approaches zero.
This behavior is consistent with the dynamics defined by the bimodal switching system, Eq.~\eqref{eqn:bimodal_switching}, which draws data from one of two states defined by two Gaussian distributions separated by a distance $\delta$.
As $\delta$ decreases, the two states become less distinct, and in the limiting case $\delta = 0$, both states are identical and variations in the switching probability $\alpha$ do not affect the dynamics.
This interplay between $\delta$ and $\alpha$ is clearly captured in the geometry of the reconstructed embedding space, Fig.~\ref{fig:goodcases}E, which collapses to a point at low $\delta$.
In the case of the Mackey--Glass system, smooth variation in parameters do not yield smooth variation in statistical properties of the generated time series.
This distorts the rectangular geometry of the sampled $(a,\tau)$ parameter space, shown in Fig.~\ref{fig:goodcases}G, due to transition-like behavior with $a$ (from a low-$a$ chaotic regime to a high-$a$ oscillatory regime).


For some datasets generated through the variation of three independent free parameters $\mathcal{X}^{(3)}$, we could detect corresponding three-dimensional structure in the estimated $(\xi_1,\xi_2,\xi_3)$ embedding.
Two selected examples are shown in Figs~\ref{fig:goodcases}I,J for datasets generated by varying:
the map parameter, $\mu \sim U(0.5,4)$, noise probability, $q \sim U(0,1)$, and noise amplitude, $b \sim U(0,3)$, in the stochastic sine map, Eq.~\eqref{eqn:stochastic_sine_map} (Fig.~\ref{fig:goodcases}I);
and varying the growth rate, $r \sim U(0.1,3)$, carrying capacity $K \sim U(0.5,5)$, and noise amplitude $\beta \sim U(0.1,10)$ in a population growth model, Eq.~\eqref{eqn:population_growth} (Fig.~\ref{fig:goodcases}J).
In both cases, the three-dimensional feature-space embedding structures the time-series dataset in a way that mirrors the three-dimensional parameter space used to generate the dataset.

These results demonstrate the ability of our method to infer low-dimensional parametric spaces as low-dimensional structure across a sufficiently rich time-series feature set for selected systems generated from models with one, two, and three dimensions of parametric freedom.

\subsection{Performance across diverse time-series datasets}

Having demonstrated that our method can successfully capture the variation of model parameters across a range of well-behaved time-series datasets, we next aimed to characterize its strengths and weaknesses across all systems, focusing on two quantitative performance metrics:
(i) the `estimated dimensionality', or the ability to infer the dimensionality of the underlying parameter space (using a heuristic to estimate $\hat{d}$ from the saturation of the residual variance $\sigma^2(d)$, cf. \textit{Methods}
); and
(ii) the `reconstruction error', the ability to generate a space $\Xi$ in which the underlying parameters can be well reconstructed (assessed as unexplained variance from a linear reconstruction $\chi$, cf. \textit{Methods}).

\begin{figure}[h]
\includegraphics[width=8cm]{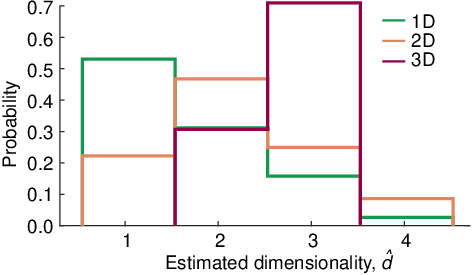}
\caption{
\textbf{The estimated dimensionality $\hat{d}$ across all systems generated from varying a single free parameter (`1D'), two free parameters (`2D'), and three free parameters (`3D').}
Probability distributions of $\hat{d}$ are shown for all datasets generated from variation of a single parameter, $m = 1$ (green),
two independent parameters, $m = 2$ (pink), and
three independent parameters, $m = 3$ (purple).
The mode of each distribution matches the true dimensionality of the underlying parameter space, $m$, but there is substantial variation.
}
\label{fig:estimDim}
\end{figure}


The distribution of estimated dimensionality $\hat{d}$ across all systems is shown in Fig.~\ref{fig:estimDim}.
While the mode of each $\hat{d}$ distribution corresponds to the true number of free parameters in the generative model, there is substantial variation.
Overestimation, $\hat{d} > m$, can occur when there is an insufficient number and sensitivity of time-series features in \textit{hctsa} to the parametric variation represented in the system.
In such cases, variation in the feature space can be dominated by noisy variation in time-series properties unrelated to parameter variation, but instead due to the stochastic generative process (including random initial conditions).
This results in a slowly saturating residual variance and hence a high estimated dimensionality.
An example is a dataset generated through one-dimensional variation in the mean offset, $s$, of a noisy sinusoid, Eq.~\eqref{eqn:noisy_shifty_sine}.
While this is a very simple type of one-dimensional variation, our feature-space method failed, estimating $\hat{d} = 4$.
The \textit{hctsa} feature set was designed with the aim of capturing subtle dynamical properties (with the majority of features operating on the $z$-scored time series), with relatively few features sensitive to simple properties like shifts in mean level (only 17 features are strongly correlated, $r > 0.9$, to $s$).
In this case, noisy variation in the thousands of other features that are unrelated to $s$, dominate the feature-space variance, leading to the substantial overestimate $\hat{d} = 4$.
In some cases, including the Lorenz oscillator $x$ and $y$, both system parameters ($r$ and $b$) drove relatively weak variation in our set of time-series features ($\rho_{95} \leq 0.7$ for $x$ and $y$ and both $r$ and $b$).
As such, both systems were estimated as $\hat{d} = 4$ despite a true dimensionality $m = 2$.

Underestimation, $\hat{d} < m$, occurred when the sampled variation in one parameter dominated the variation in time-series dynamics relative to the other sampled parameter(s).
This gives the appearance of low-dimensionality, as variation in the additional parameters resulted in relative small changes to the measured statistical properties of the dynamics.
This occurred for the `noisy shifty sine' model with two-dimensional variation in the mean-shift, $s$, and noise level, $\eta$.
As outlined above, the \textit{hctsa} feature library contains relatively few estimators of $s$ compared to $\eta$, resulting in a dominating effect of $\eta$ which overshadows the relatively subtle variations in $s$.





\begin{figure}[htb]
\includegraphics[width=\columnwidth]{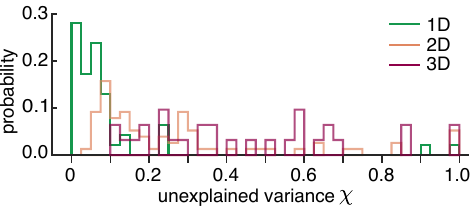}
\caption{
\textbf{Our method infers low-dimensional spaces that can reliably linearly reconstruct parameter values for most systems generated from variation of $m = 1$ or $m = 2$ free parameters, but with mixed success for the $m = 3$ case.
}
We plot the distribution of unexplained variance $\chi = 1 - R^2$, between linearly estimated parameters $\hat{p}$ from the first $\hat{d}$ embedded components and true parameters $p$.
Distributions are shown for:
46 one-parameter datasets, $\mathcal{X}^{(1)}$ (green, `1D');
74 parameters across 37 two-parameter datasets, $\mathcal{X}^{(2)}$ (orange, `2D');
and 30 parameters across 10 three-parameter datasets, $\mathcal{X}^{(3)}$ (purple, `3D').
}
\label{fig:performanceonly}
\end{figure}

We next investigated how successfully model parameters could be reconstructed as linear combinations of the $\hat{d}$ inferred dimensions $\Xi$, assessed using unexplained variance $\chi$.
Distributions of $\chi$ are shown for each parameter in each system in Fig.~\ref{fig:performanceonly}.
Performance is generally strong (with low explained variance $\chi < 0.2$ on the majority of problems) and, as expected, we generally observe the strongest reconstruction when variation across a dataset is driven by a single free parameter ($\mathcal{X}^{(1)}$), followed by the presence of a confounding parameter ($\mathcal{X}^{(2)}$), and becomes most challenging in the presence of two confounding parameters ($\mathcal{X}^{(3)}$).
Even in cases for which such an exercise may seem incredibly challenging, such as inferring the map parameter $\mu$, noise amplitude $b$, and noise probability $q$ from a stochastic sine map dataset (depicted in Fig.~\ref{fig:examplesMeasurementSpace}), our data-driven three-dimensional embedding linearly captures variation in $\mu$ ($\chi_\mu = 0.18$) and $b$ ($\chi_b = 0.22$), and contains some information about $q$ ($\chi_q = 0.63$).
We hypothesized that when there are a reasonable number of features in \textit{hctsa} that are sensitive to the variation of a given parameter, then that parameter will drive more variance across the feature space, and thus make a greater contribution to the estimated low-dimensional embedding.
This hypothesis was supported by a high observed correlation between $\chi$ and the 95th percentile of the distribution of correlations between a given parameter and the set of \textit{hctsa} features $\rho_{95}$: Pearson correlation $r(\chi,\rho_{95}) = -0.72$ (across $\mathcal{X}^{(1)}$ datasets), $r(\chi,\rho_{95}) = -0.82$ ($\mathcal{X}^{(2)}$), and $r(\chi,\rho_{95}) = -0.80$ ($\mathcal{X}^{(3)}$) (see Fig.~\ref{fig:feasibilityvsperformance} for scatters).

Taken together with the low-dimensional projection results presented above, we find that leveraging a diverse scientific literature of time-series features (as in the \textit{hctsa} feature set) allows time-series datasets to be embedded into a space that generally captures the parametric variation of the underlying model well.
The construction of the feature space (including what features are included) is crucial to the quantification of dynamical variation and in effectively weighting the relative importance of different types of dynamical variation.

\subsection{Case Study: Empirical fly data}

\begin{figure*}[htb]
\includegraphics[width=\textwidth]{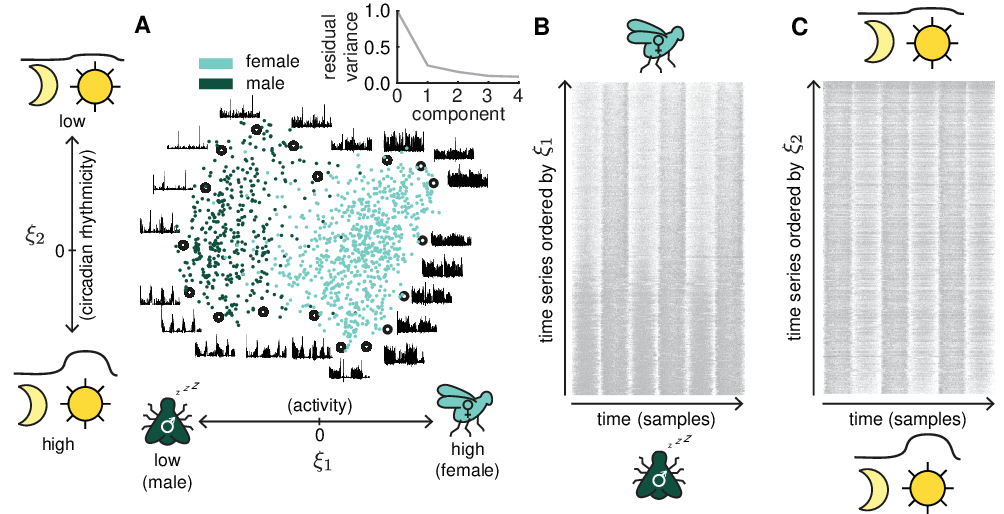}
\caption{
\textbf{Our data-driven method constructs a meaningful low-dimensional embedding space for a movement speed dataset of 1143 flies, organizing the flies by activity level and circadian rhythmicity.}
\textbf{A} The inferred data-driven embedding $(\xi_1,\xi_2)$ of the time-series dataset $\mathcal{X}$, where 12\,h time-series segments have been annotated to selected time series.
\textbf{B} Raster plot of all time series ordered by the first inferred dimension, $\xi_1$, revealing a clear variation in activity profiles of flies along this dimension.
White color corresponds to high velocity, dark color corresponds to low velocity.
To be of equal length, each time series was cut to include the first 18\,h.
\textbf{C} Raster plot of all time series ordered by the first inferred dimension $\xi_2$, revealing a visible variation of circadian rhythmicity of activity along this dimension.
}
\label{fig:flystudy}
\end{figure*}

We next aimed to investigate the performance of our feature-based dimensionality reduction approach to uncover key sources of dynamical variation underlying a real-world dataset of \textit{Drosophila melanogaster} movement.
These data were generated from a high-throughput platform for real-time tracking of animal movement \cite{Geissmann2017:EthoscopesOpenPlatforma, Jones2023:ReductionistParadigmHighthroughput}.
The dataset we analyze here is openly available \cite{geissmann_dataset} and contains time-series measurements of the movement speed of 1143 flies (408 male, 735 female) confined to an approximately one-dimensional tube and measured over the course of at least 12\,h (all $>$18\,h except 4 cases) (during which 3\,h dark and 3\,h light phases alternated successively).
The data-collection procedures are described in \citet{Geissmann2019:MostSleepDoes}.
Taking average speeds over non-overlapping 10\,s windows yielded a dataset of 1143 time series, each containing a variable number of samples, ranging from 4359--8640 (mean $=8595$).

The first two components extracted by our method $(\xi_1,\xi_2)$, had an unexplained variance of 15\%, yielding a two-dimensional projection of the time-series dataset shown in Fig.~\ref{fig:flystudy}A, where the movement time series for each fly has been colored by sex.
We first notice that the first inferred dimension $\xi_1$ separates the male and female flies, with annotations in Fig.~\ref{fig:flystudy}A revealing demonstrating that this axis acts as a measure of activity, with low-activity (mostly male) flies exhibiting large bursts of activity with an approximately 3\,h period, and high-activity (mostly female) flies exhibiting more consistent activity levels across the recording.
This is clearly visible when ordering all flies along the $\xi_1$ axis, shown as an activity raster in Fig.~\ref{fig:flystudy}B.
Consistent with this visual intuition, the time-series features that correlate most strongly with $\xi_1$ capture this variation via complexity measures (including the distributional histogram entropy \verb|EN_DistributionEntropy_hist_fd_0|, $r_\mathrm{\xi_1} = 0.97$, and measures of signal predictability like Approximate Entropy, ApEn$(1,0.1)$, $r_\mathrm{\xi_1} = 0.97$).
The second inferred dimension $\xi_2$ revealed a more subtle dynamical variation, visible most clearly from the raster plot where flies have been ordered by $\xi_2$, shown in Fig.~\ref{fig:flystudy}C.
We see that $\xi_2$ organizes flies on the strength of their circadian (day--night) rhythmicity of activity with the imposed 3\,h light--dark cycle.
Consistent with this, time-series features that most strongly correlate with $\xi_2$ are related to measures of stationarity (e.g., \verb|StatAvl250|, $r_\mathrm{\xi_2} = 0.81$) and automutual information on longer timescales (e.g., 6.7\,min: \verb|IN_AutoMutualInfoStats_40_kraskov1_4_ami40|, $r_\mathrm{\xi_2} = 0.81$).

In summary, our data-driven inference method was able to embed a large and complex dataset of fly movement time series into a behaviorally meaningful two-dimensional space in which each axis corresponds to a biologically interpretable property of the movement-speed dynamics that captures inter-fly variation across the dataset.
As per our simulated results above, this outcome is underpinned by the comprehensiveness of the \textit{hctsa} feature space; repeating the analysis using the smaller \textit{catch22} feature set \cite{Lubba:2019} was unable to capture a meaningful circadian component in $\xi_2$.

\subsection{Robustness and Sensitivity}
\label{sec:robustnsss}

\begin{figure}[htb]
\includegraphics[width=7.8cm]{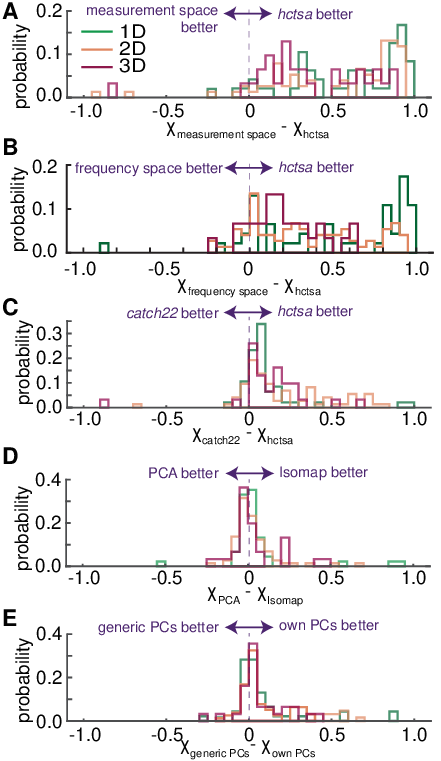}
\caption{
\textbf{Dependence of parameter reconstruction performance on methodological parameters.
}
Each panel shows the difference in unexplained variance of parameter reconstruction $\chi$ between a methodological component of our method relative to an alternative choice.
We tested the following settings:
\textbf{A} Extracting components from the raw 5000 samples of each time series (instead of the extracted \textit{hctsa} time-series feature space);
\textbf{B} Using a 100-bin Fourier power spectrum coefficient-based representation of each time series (instead of \textit{hctsa});
\textbf{C} Using the 22-feature subset, \textit{catch22}, instead of the full \textit{hctsa} feature set;
\textbf{D} Using linear PCA for dimensionality reduction (instead of the nonlinear Isomap); and
\textbf{E} Using principal components computed from a generic set of time series (`generic PCs') instead of computing them individually on each specific time-series dataset (`own PCs').
}
\label{fig:methodscomp}
\end{figure}

In this section we assess the method's strengths and limitations, including its performance with respect to the key dataset characteristics of time-series length, number of time series, and noise level.
First, across all analyzed systems, we investigated the effect of four key parameters of our methodology on reconstruction performance $\chi$:
(i) use of the \textit{hctsa} feature space rather than the raw space of time-series measurements (Fig.~\ref{fig:methodscomp}A);
(ii) use of the diverse set of \textit{hctsa} time-series features relative to a generic linear correlation-based representation (using 100-bin power spectral coefficients, including one mean-offset 0-frequency component, Fig.~\ref{fig:methodscomp}B);
(iii) use of \textit{hctsa} rather than the compact reduced subset of 22 features, \textit{catch22} (Fig.~\ref{fig:methodscomp}C);
(iv) use of the nonlinear dimension-reduction method, Isomap \cite{Tenenbaum2000:GlobalGeometricFramework} rather than the linear dimension-reduction method, PCA (Fig.~\ref{fig:methodscomp}D); and
(v) use of PCs from a generic set of time-series data rather than computing the PCs specifically on the dataset of interest (Fig.~\ref{fig:methodscomp}E).
In each case shown in Fig.~\ref{fig:methodscomp}, the two alternative methodological choices are compared as the difference in $\chi$ between the method we present and an alternative choice; higher values ($>0$) indicate better performance of the method we present than the alternative.

Our first result, in Fig.~\ref{fig:methodscomp}A, demonstrates that the strong performance of our method is overwhelmingly driven by the feature-space embedding, with vastly reduced average performance when dimension reduction is instead performed instead in the space of the raw (ordered) time-series measurements (labeled $\chi_\mathrm{measurement\ space}$).
Parameters could only be well-reconstructed in the raw measurement space for problems in which the parameters caused clear changes in the overall shape of time series (such as the slope $m$ in the `noisy trendy sine' system, or the mean shift $s$ in `noisy shifty sine' system).
The usefulness of a comprehensive time-series feature library to capture diverse sources of variation in statistical properties of time series is demonstrated in the comparisons to two alternative feature sets: Fourier power spectral coefficients (Fig.~\ref{fig:methodscomp}B) and one based on the representative subset of 22 features, \textit{catch22}~\cite{Lubba:2019} (Fig.~\ref{fig:methodscomp}C).
Both of these feature sets are much more restricted in the types of statistical properties they can capture and resulted in substantial reductions in performance for the vast majority of problems.
While using Isomap for dimension reduction yielded better performance on average than using PCA, for most problems the performance was similar (Fig.~\ref{fig:methodscomp}D), with the nonlinearity of Isomap being most important for $\mathcal{X}^{(1)}$ systems, but less crucial for $\mathcal{X}^{(2)}$ and $\mathcal{X}^{(3)}$ datasets.
Finally we compare generic low-dimensional components, trained on a diverse dataset of 1000 time series (the \textit{Empirical1000} set \cite{Fulcher2020:SelforganizingLivingLibrary}) from across science (labeled $\chi_\mathrm{generic PCs}$ in Fig.~\ref{fig:methodscomp}E), to those computed directly from the analyzed time-series dataset (labeled $\chi_\mathrm{own PCs}$ in Fig.~\ref{fig:methodscomp}E).
Our results demonstrate a general improvement from tailoring the low-dimensional components to a given dataset, but on many problems the difference is relatively small.
This points to common, generic and informative components that can be derived from the behavior of \textit{hctsa} on general time-series datasets that capture the types of statistical time-series properties that capture the statistical effect of parameter variation on many of the systems studied here.

\begin{figure}[h]
\includegraphics[width = 8cm]{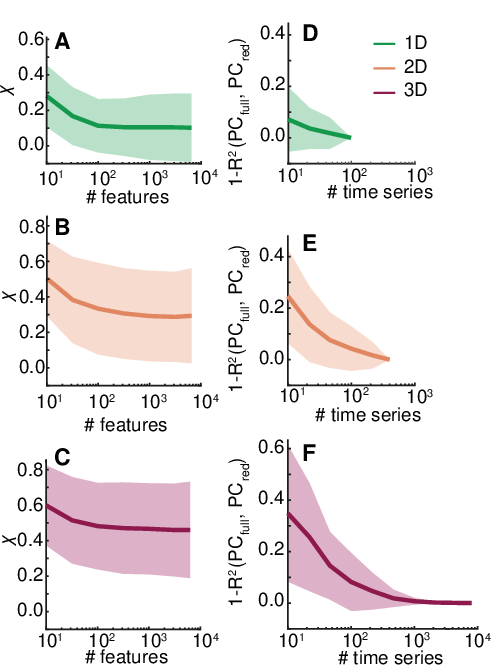}
\caption{
\textbf{Dependence of our feature-based dimension reduction method's performance on (randomly selected) subsets of features, and with the number of time series included in the dataset.
}
\textbf{A} Unexplained variance of parameter reconstruction $\chi$ using a given number of randomly sampled subset of \textit{hctsa} features.
The line shows the mean across 10 repetitions of random sampling; shading indicates the standard deviation.
\textbf{B} Difference between extracted principal component dimensions obtained with a full dataset (labeled PC$_\mathrm{full}$) versus those obtained using a random subset of time series from the dataset (labeled PC$_\mathrm{red}$).
}
\label{fig:robustness}
\end{figure}


We next aimed to investigate whether similar reconstruction performance (indexed via $\chi$) could be obtained with random feature subsets from \textit{hctsa} (to potentially reduce the computational cost of computing all $>7000$ features), and how performance scales with the number of time series (to understand the feasibility of our approach in relatively data-poor settings).
Unexplained variance is shown as a function of the number of randomly selected features in Figs~\ref{fig:robustness}A, B, C (for 1-dimensional, 2-dimensional, and 3-dimensional systems, respectively).
We found a plateau in performance at about 300 features for all system dimensionalities, suggesting that there is clear scope for performing well with feature sets of substantially reduced size than the full \textit{hctsa} set (particularly in future work if more principled feature-selection methods were used over the simple random sampling approach used in this test).


To investigate the effect of dataset size on performance, we measured how similar the principal components extracted from the full dataset (PC$_\mathrm{full}$) are to those extracted from a random subset of data (PC$_\mathrm{red}$) are, as $1 - R^2(\mathrm{PC}_\mathrm{full},\mathrm{PC}_\mathrm{red})$.
A value of this metric near zero reflects that the reduced feature components computed using the subset of time series is strongly correlated to that obtained with the full set.
Results shown in Figs~\ref{fig:robustness}D,E,F for $\mathcal{X}^{(1)}$, $\mathcal{X}^{(2)}$, and $\mathcal{X}^{(3)}$ datasets, respectively.
For the 1- and 2-dimensional cases, which were limited to a maximum of 100 and 400 time series, respectively, we did not observe clear plateau behavior, but saturation was observed for the 3-dimensional systems (at about 1000 time series).
In all cases, relatively small time-series subsets yielded highly correlated approximations to the low-dimensional components computed using the full dataset, indicating the applicability of our method in more data-constrained settings than those demonstrated in our simulations here.

\section{Discussion}


We have introduced an unsupervised, data-driven method for estimating dominant sources of statistical variation across a time-series dataset using a library of time-series features.
Our key hypothesis was that the \emph{hctsa} library of time-series statistics was diverse enough to allow interpretable inference of low-dimensional structure for a wide range of dynamical systems.
We found broad support for our hypothesis across a range of simulated linear and nonlinear dynamical systems in discrete and continuous time, where it often extracted low-dimensional representations of the time-series dataset that matched the underlying parametric degrees of freedom in the (unknown) generative model.
Our results open the prospect of interpretable estimation of relevant parameters of dynamical systems \emph{without prior knowledge of the generating model} that, since our approach is grounded in algorithms derived from time-series theory, can provide deeper understanding of the structure of a given time-series dataset.
We include a demonstration of the method on a dataset of fly movement time series, yielding a biologically meaningful representation of the dataset in terms of sex and circadian rhythmicity of the flies.
Our approach is generally applicable to finding structure in large, complex time-series datasets---which are increasingly common across science and industry---and could be adapted to many settings, including tracking sources of non-stationarity across shorter windows of a long time series~\cite{Guttler2001:ReconstructionParameterSpaces,Owens2024:ParameterInferenceNonstationarya}.

Relative to prior work, that has required manual identification of appropriately sensitive summary statistics of the underlying statistical variation across a dataset~\cite{Guttler2001:ReconstructionParameterSpaces}, a key innovation of this work is the ability to infer the relevant statistics of dynamical variation directly from data using a comprehensive library of summary statistics.
This allows us to detect many different sources of possible statistical variation across the dataset (from variation in distributional shape to complex linear and nonlinear self-correlation structures), which appears to be the key driver of the methods performance relative to alternative feature sets like Fourier spectral components (Fig.~\ref{fig:methodscomp}B) or the \textit{catch22} subset~\cite{Lubba:2019} (Fig.~\ref{fig:methodscomp}C), although there is clear scope for feature-set reduction (Fig.~\ref{fig:robustness}) in future, which would bring substantial computational benefits.
Converting the dimension-reduction problem from one in the space of the time-series measurements themselves (`signal space') \cite{Ashraf2023:SurveyDimensionalityReduction} to a set of statistical properties has advantages beyond its dramatically improved performance (Fig.~\ref{fig:methodscomp}A).
And, although the simulations performed here consider sets of time series of equal length, our method can be applied to sets of time series of unequal length, since all time series are mapped to a feature vector of a common dimensionality.

From the perspective of likelihood-free inference methods such as Approximate Bayesian Computation (ABC) \cite{Toni2009:ApproximateBayesianComputation, Beaumont2010:ApproximateBayesianComputation, Sisson2018:OverviewABC}, the success of inference depends critically on the availability of informative summary statistics.
There is a research direction to attempt to learn these features \cite{Fearnhead2012:ConstructingSummaryStatistics}.
Our results suggest that the \textit{hctsa} feature library already provides a rich basis of candidate summary statistics for low-dimensional dynamical parameter inference, often containing multiple features strongly sensitive to each parameter across a diverse set of simulated systems (Figs~\ref{fig:feasibilityonly}, \ref{fig:performanceonly}).

Although the simulated systems studied here were explicitly constructed with a small number of free parameters, a substantial body of work on ``sloppy models'' suggests that dynamical systems with apparently high-dimensional parameter spaces often exhibit low effective dimensionality in their observable behavior, with dynamics governed by a small number of stiff parameter combinations~\cite{Brown2003:StatisticalMechanicalApproaches, Gutenkunst2007:UniversallySloppyParameter}.
This provides a theoretical rationale for expecting low-dimensional structure in feature space to persist beyond the explicitly low-dimensional examples considered here.


Notably, the specific choices of feature selection, dimensionality reduction, and manifold learning algorithm are less critical in our setting; as demonstrated in Fig.~\ref{fig:methodscomp}, performance is driven primarily by the breadth and diversity of the feature library, with only modest differences between linear and nonlinear embeddings once an adequately expressive representation is in place.
We can expect that refining the inference part of our work, e.g., using alternative approaches to manifold learning, will certainly improve the results but the strength of our hypothesis rests on the diversity of the feature library.
While representations could be learned directly from raw time series using neural architectures such as autoencoders, our approach deliberately relies on a library of scientifically interpretable features allowing low-dimensional structure to be related to established dynamical properties.
We note that, to detect an underlying shared explanatory dimension of variation, the feature set must contain \textit{multiple} statistical features that are all sensitive to the statistical variation induced by the underlying parametric freedom.
To take a simple example of a dataset with a single source of variation across time series---the time-series length---then our method would struggle to detect this variation as low-dimensional structure, unless the feature set contained a sufficient number of length-dependent features (the inter-correlation of which on datasets for which there is variation in length would yield resulting low-dimensional structure).
A trade-off is involved in making the feature set sufficiently comprehensive to contain multiple estimators of any given source of statistical variation (that we would like our method to be sensitive to), but while noting that its makeup can also bias the extracted dimensions to those that reflect the make-up of the feature set more than the specific low-dimensional structure of any given dataset.
Although dataset-tailored components often yield substantial improvements over computing low-dimensional components from a diverse generic set of real-world and simulated time series~\cite{Fulcher2020:SelforganizingLivingLibrary}
(Fig.~\ref{fig:methodscomp}E), the generic components often provide a reasonable approximation to the tailored components, suggesting potential for future work to better overcome the bias in the extracted components resulting from the specific make-up of any time-series feature set.

The mapping from our extracted dimensions to underlying parameters works most clearly in one-dimensional systems but, because our method detects unique sources of statistical variation across a dataset, its success relies on the parameters, in the case of multiple parameters, shaping the dynamics in distinct and detectable ways.
It is commonly the case, of course, that model parameters are not identifiable from data, even when the exact model for the data is known: in this case our approach will necessarily fail.
Consider, for example, the bimodal switching system we analyze here, Eq.~\eqref{eqn:bimodal_switching}, where the switching-rate parameter $\alpha$ has a clear effect on the dynamics when the two states are distinct (at high state separation $\delta$), but has no effect when the two states are identical (at $\delta = 0$).
Our embedding accurately reflects this dependence of the parameters on the dynamical structure, with the component correlated to $\alpha$ reducing to a point at one extreme of the component correlated to $\delta$, and branching out (Fig.~\ref{fig:goodcases}E).
This effect complicates the inference of datasets generated from higher-dimensional parametric freedoms, where there is the additional complication of some more nuanced parameters (or those with relatively `low' variability) having relatively small effects on the resulting features extracted from them, which can often be drowned out by the dominant role of more obvious parameters with larger and more salient effects on the dynamics (like those driving changes in low-order trends, or strong frequency components, which often obscure the secondary effects of other parameters).
Therefore, while the dominant sources of variation in statistical properties across a time-series dataset need not correspond to the free parameters of the underlying generative model for a time-series dataset, we show many cases here where these inferred components strongly recapitulate the underlying parametric freedoms.
An understanding of the number of free parameters that underlie statistical variation across a collection of time series, and the types of dynamical properties they control, can be crucial in guiding the development of an explanatory statistical model for the data.

In summary, we have demonstrated the existence of a sufficiently diverse library of interpretable time-series features to allow a straightforward combination of feature extraction and dimension reduction and can be used to estimate and interpret the effective parametric dimensionality of dynamical systems directly from data, even when the governing model is unknown.
In the future, the basic theoretical framework and methodologies developed in this work may form the basis for a powerful way of linking the theoretical formulation of statistical time-series models with the empirical practice of collecting and analyzing time series.
Given the numerous time-series datasets emerging for real-world problems, the low-dimensional representations presented here can provide interpretable understanding of the potential generative mechanisms underlying apparent complexity.


\begin{acknowledgments}
S.R.S. acknowledges support from EPSRC EP/L016737/1.
B.D.F. acknowledges support from the Australian Research Council (FT240100418).
The authors would like to thank Kieran Owens and Brendan Harris for helpful feedback on the manuscript.
\end{acknowledgments}

\newpage
\appendix

\section{Systems}

\setcounter{table}{0}
\renewcommand{\thetable}{A\arabic{table}}%
\setcounter{figure}{0}
\renewcommand{\thefigure}{A\arabic{figure}}%



\begin{figure}[htb]
\includegraphics[width=7cm]{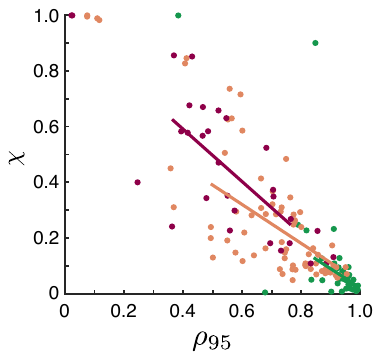}
\caption{
\textbf{When a greater proportion of features are sensitive to the variation of a given parameter, that parameter is recovered more accurately by our method.}
We show a scatter plot of reconstruction performance, $\chi$ [as in Fig.~\ref{fig:performanceonly} as a function of the feasibility, $\rho_{95}$ (as in Fig.~\ref{fig:feasibilityonly}).
Correlations are $|\rho| = 0.72$ (1-dim), $|\rho| = 0.82$ (2-dim), and $|\rho| = 0.80$ (3-dim)].
}
\label{fig:feasibilityvsperformance}
\end{figure}

\subsection{Time-Series Datasets}
\label{sec:all_systems}

This section describes details of each simulated system.
Systems was simulated using Matlab code from \url{https://github.com/benfulcher/TimeSeriesGeneration/}~\cite{Zenodo_TimeSeriesGeneration}.

\subsection{Stochastic Systems}

\subsubsection{Noisy Trendy Sine}

This simple time-series system generates a sinusoid with period $T$, with a linear trend of gradient $m$ and additive noise of standard deviation $\eta$.
Time series, $x_t$, are generated according to the following model:

\begin{equation}
\label{eqn:noisy_trendy_sine}
x_t = \sin(2\pi t/T) + m t/N + \eta n_t\,,
\end{equation}
for a period $T$, gradient $m$, and noise standard deviation $\eta$, and $n_t \sim \mathcal{N}(0,1)$.
In generating datasets, we varied $T$, $m$, and $\eta$ as in the table below:

\begin{table}[h]
\centering
\begin{tabular}{cc}
Parameter & Values\\ \hline
$T$         & $U(10,100)$, or 30.\\
$m$         & $U(-5,5)$, or 0.\\
$\eta$       & $U(0,3)$, or 0.
\end{tabular}
\end{table}

\subsubsection{Noisy Shifty Sine}

This model generates a noisy sinusoid with a constant mean offset:
\begin{equation}
\label{eqn:noisy_shifty_sine}
x_t = \sin(2\pi t/T) + s + \eta n_t\,,
\end{equation}
for period $T$, mean offset $s$, and noise standard deviation $\eta$, and $n_t \sim \mathcal{N}(0,1)$.

\begin{table}[h]
\centering
\begin{tabular}{cc}
Parameter & Values\\ \hline
$T$         & $U(10,100)$ or $30$.\\
$\beta$	  & $U(-5,5)$ or $0$.\\
$\eta$    & $U(0,3)$ or $0$.
\end{tabular}
\end{table}

\subsubsection{Autoregressive (AR) process}

We generated data from an autoregressive process \cite{Chatfield04} with amplitude $\alpha$, memory $\tau$, and a constant Gaussian noise term $\eta$.

\begin{equation}
x_t = \alpha \frac{1}{\tau}\sum_{i=1}^\tau x_{t-i} + \eta_t\,,
\end{equation}
where $\eta_t \sim \mathcal{N}(0,1)$ are i.i.d. samples from a Gaussian distribution. Initial $\tau$ values of $x$ are also $\sim \mathcal{N}(0,1)$.

\begin{table}[h]
\centering
\begin{tabular}{cc}
Parameter & Values\\ \hline
$\alpha$ & $U(0.8, 0.999)$ or $0.9$.\\
$\tau$ & $\{2,3,4,5,6,7,8,9,10\}$ or $2$.\\
\end{tabular}
\end{table}

\subsubsection{Bimodal switching model}

This two-state model samples from one of two Gaussian distributions, each with a different mean \cite{Fulcher2012}.
At each time point, there is a probability of switching between the two states:

\begin{align}
\begin{split}
\label{eqn:bimodal_switching}
x_t &= n_t + v_t \delta\,, \\
{v_t} &= Y_t(\alpha) v_{t-1} + (1 - Y_t(\alpha))(1 - v_{t-1}) \,,
\end{split}
\end{align}
for a time series $x_t$, underlying state variable $v_t$, $n_t \sim \mathcal{N}(0,1)$.
State switching is controlled by the Bernoulli random variable, $Y_t$, which takes a value of 1 with probability $\alpha$, and a value of 0 with probability $1 - \alpha$.

\begin{table}[h]
\centering
\begin{tabular}{cc}
Parameter & Values\\ \hline
$\delta$	& $U(0,6)$ or $3$.\\
$\alpha$	& $U(0,1)$ or $0.5$.
\end{tabular}
\end{table}

\subsubsection{Population growth}

The stochastic population growth model, with a limited carrying capacity and a variable growth rate \cite{Levins1969}, is governed by the following equation:
\begin{equation}
\label{eqn:population_growth}
\dot x = r x (K - x) + x \eta_\beta\,,
\end{equation}
with maximum population size, $K$, growth rate, $r$, and the amplitude of an i.i.d. noise term $\eta_\beta$, drawn from a uniform distribution with a range from 0 to $\beta$.

\begin{table}[h]
\centering
\begin{tabular}{cc}
Parameter & Values\\ \hline
$r$	& (0.1, 3) or 3 \\
$K$ & (0.5, 5) or 5 \\
$\beta$ & (0.1, 10) or 0.1
\end{tabular}
\end{table}

\subsubsection{Stochastic sine map}

The stochastic sine map model was introduced by \citet{Freitas2009:FailureDistinguishingColored} and is defined as:
\begin{equation}
\label{eqn:stochastic_sine_map}
x_{t+1} = \mu \sin(x_t) + Y_t(q)\eta_t(b)\,,
\end{equation}
for a sinusoidal amplitude $\mu$,
a Bernoulli random variable $Y$ ($Y = 0$ with probability $q$, and $Y = 1$ with probability $1 - q$),
and an identically and independently distributed random variable $\eta_t$ which has a uniform distribution between $-b$ and $b$. Initial $x_0$ is set according to a uniform random distribution between $-1$ and $1$.

\begin{table}[h]
\centering
\begin{tabular}{cc}
Parameter & Values\\ \hline
$\mu$ & $U(0.5,4)$ or $2.4$.\\
$q$	& $U(0,1)$ or $0.5$.\\
$b$ & $U(0,3)$ or $1$.
\end{tabular}
\end{table}

\subsection{Deterministic Flows}

Here we use the term `flow' to describe a dynamical system formulated in continuous time.
All flows were simulated in Matlab using the ordinary differential equation solver \texttt{ode45}.
Each system was then evaluated on an even time grid of an appropriate resolution for each system (listed below for each system).
In all cases, the first 500 samples were considered a transient and removed, to minimize the dependence on initial conditions.

\subsubsection{Van der Pol Oscillator}

The van der Pol oscillator is a deterministic nonlinear flow with negative dampening for low amplitudes of $x$ and positive dampening for high values \cite{VanderPol1926, Sprott03}.
The dynamics are governed by
\begin{equation}
\label{eqn:vanderPol}
\ddot x - c (1 - x^2) \dot x + k x = 0\,,
\end{equation}
where $c$ controls the degree of dampening (positive or negative) and $k$ affects the oscillation frequency.
The system was evaluated at a temporal sampling rate of 1/6\,Hz.

\begin{table}[h]
\centering
\begin{tabular}{cc}
Parameter & Values\\ \hline
$c$	& $U(0.1,5)$ or 1.\\ 
$k$ & $U(0.1,5)$ or 1. 
\end{tabular}
\end{table}


\subsubsection{R\"ossler Oscillator}

A three-dimensional chaotic flow proposed by R\"ossler in 1976 \citet{Rossler1976, Sprott03}.
It is one of the simplest chaotic flows with a single quadratic nonlinearity ($zx$).
The dynamics in the three-dimensional phase space are governed by
\begin{eqnarray}
\begin{split}
\label{eqn:rossler}
    \dot x &=& - y - z \,, \\
    \dot y &=& x + a y \,, \\
    \dot z &=&  b + z (x - c)\,,
\end{split}
\end{eqnarray}
for parameters $a$, $b$, and $c$.
The sampling rate was set to 1/5\,Hz.

\begin{table}[h]
\centering
\begin{tabular}{cc}
Parameter & Values\\ \hline
$a$	& (0.1, 0.3) or 0.2 \\ 
$b$ & (0.01, 2) or 0.2 \\ 
$c$ & (4.5, 20) or 5.7 
\end{tabular}
\end{table}

\subsubsection{Lorenz Attractor}

A three-dimensional chaotic flow introduced by Lorenz \cite{Lorenz1963, Sprott03} while studying atmospheric convection.
Dynamics in the three-dimensional phase space, $(x,y,z)$, are governed by the following equations:
\begin{align}
\begin{split}
\label{eqn:lorenz}
{\dot x} &= s(y - x)\,,\\
{\dot y} &= x(r - z) - y\,,\\
{\dot z} &= xy - bz\,,
\end{split}
\end{align}
with free parameters $s$, $r$, and $b$.
The sampling rate was set to 1/5\,Hz.

\begin{table}[h]
\centering
\begin{tabular}{cc}
Parameter & Values\\ \hline
$s$	& $U(8,30)$ or $10$.\\
$r$	& $U(35,60)$ or $35$.\\
$b$	& $U(1,2.8)$ or $2.6667$.
\end{tabular}
\end{table}


\subsubsection{Mackey--Glass system}

The Mackey--Glass system is a delay differential equation corresponding to high-dimensional chaotic flow \cite{Mackey77, Glass1988}.

\begin{align}
\label{eqn:mackey_glass}
\frac{dx(t)}{dt} = \frac{ax(t - \tau)}{1 + x(t - \tau)^{10}} - 0.1x(t)\,,
\end{align}
for parameters $a$ and $\tau$.

\begin{table}[h]
\centering
\begin{tabular}{cc}
Parameter & Values\\ \hline
$a$	& $U(0.15, 1.5)$ or 1.\\
$\tau$	& $\{10, 11, ..., 39, 40\}$ or 17.
\end{tabular}
\end{table}

\subsection{Deterministic Maps}

Here we refer to a `map' as a system with dynamics formulated as an iterative relationship in discrete time.

\subsubsection{Logistic Map}

The Logistic map is a simple one-dimensional map that can exhibit chaotic dynamics, and is indeed a paradigmatic example of the phenomenon \cite{May1976, Sprott03}.
Dynamics are governed by:
\begin{equation}
x_{t+1} = a x_t (1 - x_t)\,,
\end{equation}
where $a$ is a parameter.
%
For the parameter range used here, $a \sim U(3.5, 4)$, both periodic and chaotic dynamics can occur.

\subsubsection{Predator--prey system}

Here we consider a discrete-time formulation of a predator--prey system that models the populations of an interacting predator and a prey species, $x$ and $y$ \cite{May1972, Hoppensteadt:2006}.
The $(x,y)$ population dynamics are governed by:
\begin{align}
\begin{split}
\label{eqn:predator_prey}
x_t &= x_{t-1} \exp[r (1 - x_{t-1}/K)] - \alpha  y_{t-1}\,, \\
y_t &= x_{t-1} [1 - \exp (-\alpha y_{t-1})]\,,
\end{split}
\end{align}
for parameters $r$, $K$, and $\alpha$.

\begin{table}[h]
\centering
\begin{tabular}{cc}
Parameter & Values\\ \hline
$r$	& (0.3, 1) or 0.5 \\
$K$ & (0.7, 1.2) or 1 \\
$\alpha$ & 5
\end{tabular}
\end{table}


\subsection{Other}

\subsubsection{Self-affine}

Time series with an power-law characteristic in the power spectrum with scaling exponent $\alpha$.
The time series are generated by an inverse Fourier transform (IFFT) from the desired power-law spectrum with random phase \cite{Fox1987} using Matlab.

\begin{table}[h]
\centering
\begin{tabular}{cc}
Parameter & Values\\ \hline
$\alpha$	& (-1, 3)
\end{tabular}
\end{table}



\begin{thebibliography}{40}
\providecommand{\natexlab}[1]{#1}

\bibitem[{Hastie et~al.(2009)Hastie, Tibshirani, Friedman, and
  Friedman}]{Hastie2009:ElementsStatisticalLearning}
T.~Hastie, R.~Tibshirani, J.~H. Friedman, and J.~H. Friedman.
\newblock \emph{The Elements of Statistical Learning: Data Mining, Inference,
  and Prediction}, volume~2.
\newblock Springer (2009).

\bibitem[{Owens and Fulcher(2025)}]{Owens2025:TimeseriesDimensionReduction}
K.~S. Owens and B.~D. Fulcher.
\newblock Time-series dimension reduction: A comprehensive review and
  conceptual unification of algorithms  (2025).

\bibitem[{Takens(1981)}]{Takens81}
F.~Takens.
\newblock Detecting strange attractors in turbulence.
\newblock \emph{Lect. Notes Math.} \textbf{898}, 366 (1981).

\bibitem[{Kantz and Schreiber(2004)}]{Kantz04}
H.~Kantz and T.~Schreiber.
\newblock \emph{Nonlinear Time Series Analysis}.
\newblock Cambridge University Press, Cambridge, 2nd edition (2004).

\bibitem[{Owens and Fulcher(2024)}]{Owens2024:ParameterInferenceNonstationarya}
K.~S. Owens and B.~D. Fulcher.
\newblock Parameter inference from a non-stationary unknown process.
\newblock \emph{Chaos: An Interdisciplinary Journal of Nonlinear Science}
  \textbf{34}, 101501 (2024).

\bibitem[{G{\"u}ttler et~al.(2001)G{\"u}ttler, Kantz, and
  Olbrich}]{Guttler2001:ReconstructionParameterSpaces}
S.~G{\"u}ttler, H.~Kantz, and E.~Olbrich.
\newblock Reconstruction of the parameter spaces of dynamical systems.
\newblock \emph{Physical Review E} \textbf{63}, 056215 (2001).

\bibitem[{Fulcher et~al.(2013)Fulcher, Little, and
  Jones}]{Fulcher2013:HighlyComparativeTimeseries}
B.~D. Fulcher, M.~A. Little, and N.~S. Jones.
\newblock Highly comparative time-series analysis: The empirical structure of
  time series and their methods.
\newblock \emph{Journal of The Royal Society Interface} \textbf{10}, 20130048
  (2013).

\bibitem[{Fulcher and Jones(2017)}]{Fulcher2017:HctsaComputationalFramework}
B.~D. Fulcher and N.~S. Jones.
\newblock {\emph{Hctsa}}: {{A}} computational framework for automated
  time-series phenotyping using massive feature extraction.
\newblock \emph{Cell Systems} \textbf{5}, 527 (2017).

\bibitem[{Tenenbaum et~al.(2000)Tenenbaum, de~Silva, and
  Langford}]{Tenenbaum2000:GlobalGeometricFramework}
J.~B. Tenenbaum, V.~de~Silva, and J.~C. Langford.
\newblock A global geometric framework for nonlinear dimensionality reduction.
\newblock \emph{Science} \textbf{290}, 2319 (2000).

\bibitem[{Chatfield(2004)}]{Chatfield04}
C.~Chatfield.
\newblock \emph{The Analysis of Time Series}.
\newblock CRC Press LLC (2004).

\bibitem[{May(1972)}]{May1972}
R.~M. May.
\newblock {Limit cycles in predator-prey communities}.
\newblock \emph{Science} \textbf{177}, 900 (1972).

\bibitem[{Verhulst(1845)}]{verhulst1845loi}
P.~F. Verhulst.
\newblock La loi d’accroissement de la population.
\newblock \emph{Nouv. Mem. Acad. Roy. Soc. Belle-lettr. Bruxelles} \textbf{18}
  (1845).

\bibitem[{van~der Pol(1926)}]{VanderPol1926}
B.~van~der Pol.
\newblock {On “relaxation-oscillations”}.
\newblock \emph{Philosophical Magazine and Journal of Science} \textbf{2}, 978
  (1926).

\bibitem[{Freitas et~al.(2009)Freitas, Letellier, and
  Aguirre}]{Freitas2009:FailureDistinguishingColored}
U.~S. Freitas, C.~Letellier, and L.~A. Aguirre.
\newblock Failure in distinguishing colored noise from chaos using the ``noise
  titration'' technique.
\newblock \emph{Physical Review E} \textbf{79}, 035201 (2009).

\bibitem[{May(1976)}]{May1976}
R.~M. May.
\newblock {Simple mathematical models with very complicated dynamics}.
\newblock \emph{Nature} \textbf{261}, 459 (1976).

\bibitem[{Lorenz(1963)}]{Lorenz1963}
E.~N. Lorenz.
\newblock {Deterministic nonperiodic flow}.
\newblock \emph{Journal of the atmospheric sciences} \textbf{20}, 130 (1963).

\bibitem[{R{\"{o}}ssler(1976)}]{Rossler1976}
O.~E. R{\"{o}}ssler.
\newblock {An equation for continuous chaos}.
\newblock \emph{Physics Letters A} \textbf{57}, 397 (1976).

\bibitem[{Mackey and
  Glass(1977{\natexlab{a}})}]{Mackey1977:OscillationChaosPhysiological}
{\relax MC}.~Mackey and L.~Glass.
\newblock Oscillation and chaos in physiological control systems.
\newblock \emph{Science} \textbf{197}, 287 (1977{\natexlab{a}}).

\bibitem[{Malamud and Turcotte(1999)}]{malamud1999self}
B.~D. Malamud and D.~L. Turcotte.
\newblock Self-affine time series: I. generation and analyses.
\newblock In \emph{Advances in Geophysics}, volume~40, pp. 1--90. Elsevier
  (1999).

\bibitem[{Hoppensteadt(2006)}]{Hoppensteadt:2006}
F.~Hoppensteadt.
\newblock {P}redator-prey model.
\newblock \emph{Scholarpedia} \textbf{1}, 1563 (2006).

\bibitem[{Lubba et~al.(2019)Lubba, Sethi, Knaute et~al.}]{Lubba:2019}
C.~H. Lubba, S.~S. Sethi, P.~Knaute, et~al.
\newblock {catch22: CAnonical Time-series CHaracteristics}.
\newblock \emph{Data Mining and Knowledge Discovery} pp. 1--32 (2019).

\bibitem[{Fulcher(2012)}]{Fulcher2012}
B.~D. Fulcher.
\newblock \emph{{Highly Comparative Time-Series Analysis}}.
\newblock Ph.D. thesis, University of Oxford (2012).

\bibitem[{Glass et~al.(1988)Glass, Mackey, and Zweifel}]{Glass1988}
L.~Glass, M.~C. Mackey, and P.~F. Zweifel.
\newblock \emph{{From Clocks to Chaos: The Rhythms of Life}}.
\newblock Princeton University Press (1988).

\bibitem[{Levins(1969)}]{Levins1969}
R.~Levins.
\newblock {The effect of random variations of different types on population
  growth.}
\newblock \emph{Proceedings of the National Academy of Sciences of the United
  States of America} \textbf{62}, 1061 (1969).

\bibitem[{Mackey and Glass(1977{\natexlab{b}})}]{Mackey77}
M.~C. Mackey and L.~Glass.
\newblock Oscillation and chaos in physiological control systems.
\newblock \emph{Science} \textbf{197}, 287 (1977{\natexlab{b}}).

\bibitem[{Sprott(2003)}]{Sprott03}
J.~C. Sprott.
\newblock \emph{Chaos and Time-Series Analysis}.
\newblock Oxford University Press, New York (2003).

\bibitem[{Geissmann et~al.(2017)Geissmann, Rodriguez, Beckwith
  et~al.}]{Geissmann2017:EthoscopesOpenPlatforma}
Q.~Geissmann, L.~G. Rodriguez, E.~J. Beckwith, et~al.
\newblock Ethoscopes: {{An}} open platform for high-throughput ethomics.
\newblock \emph{PLOS Biology} \textbf{15}, e2003026 (2017).

\bibitem[{Jones et~al.(2023)Jones, Willis, Firth, Giachello, and
  Gilestro}]{Jones2023:ReductionistParadigmHighthroughput}
H.~Jones, J.~A. Willis, L.~C. Firth, C.~N. Giachello, and G.~F. Gilestro.
\newblock A reductionist paradigm for high-throughput behavioural
  fingerprinting in {{Drosophila}} melanogaster.
\newblock \emph{eLife} \textbf{12}, RP86695 (2023).

\bibitem[{Geissmann et~al.(2026)Geissmann, Beckwith, and
  Gilestro}]{geissmann_dataset}
Q.~Geissmann, E.~J. Beckwith, and G.~F. Gilestro.
\newblock Raw ethoscope recordings — geissmann, beckwith \& gilestro 2019
  (sci. adv.) — data set 1 of 2 (20160404\_overnight\_dsd) (2026).

\bibitem[{Geissmann et~al.(2019)Geissmann, Beckwith, and
  Gilestro}]{Geissmann2019:MostSleepDoes}
Q.~Geissmann, E.~J. Beckwith, and G.~F. Gilestro.
\newblock Most sleep does not serve a vital function: {{Evidence}} from
  {{Drosophila}} melanogaster.
\newblock \emph{Science Advances} \textbf{5}, eaau9253 (2019).

\bibitem[{Fulcher et~al.(2020)Fulcher, Lubba, Sethi, and
  Jones}]{Fulcher2020:SelforganizingLivingLibrary}
B.~D. Fulcher, C.~H. Lubba, S.~S. Sethi, and N.~S. Jones.
\newblock A self-organizing, living library of time-series data.
\newblock \emph{Scientific Data} \textbf{7}, 213 (2020).

\bibitem[{Ashraf et~al.(2023)Ashraf, Anowar, Setu
  et~al.}]{Ashraf2023:SurveyDimensionalityReduction}
M.~Ashraf, F.~Anowar, J.~H. Setu, et~al.
\newblock A {{Survey}} on {{Dimensionality Reduction Techniques}} for
  {{Time-Series Data}} \textbf{11}, 42909 (2023).

\bibitem[{Toni et~al.(2009)Toni, Welch, Strelkowa, Ipsen, and
  Stumpf}]{Toni2009:ApproximateBayesianComputation}
T.~Toni, D.~Welch, N.~Strelkowa, A.~Ipsen, and M.~P.~H. Stumpf.
\newblock Approximate {{Bayesian}} computation scheme for parameter inference
  and model selection in dynamical systems.
\newblock \emph{Journal of The Royal Society Interface} \textbf{6}, 187 (2009).

\bibitem[{Beaumont(2010)}]{Beaumont2010:ApproximateBayesianComputation}
M.~A. Beaumont.
\newblock Approximate {{Bayesian Computation}} in {{Evolution}} and
  {{Ecology}}.
\newblock \emph{Annual Review of Ecology, Evolution, and Systematics}
  \textbf{41}, 379 (2010).

\bibitem[{Sisson et~al.(2018)Sisson, Fan, and
  Beaumont}]{Sisson2018:OverviewABC}
S.~A. Sisson, Y.~Fan, and M.~A. Beaumont.
\newblock Overview of {{ABC}}.
\newblock In \emph{Handbook of {{Approximate Bayesian Computation}}}. {Chapman
  and Hall/CRC} (2018).

\bibitem[{Fearnhead and
  Prangle(2012)}]{Fearnhead2012:ConstructingSummaryStatistics}
P.~Fearnhead and D.~Prangle.
\newblock Constructing summary statistics for approximate {{Bayesian}}
  computation: Semi-automatic approximate {{Bayesian}} computation.
\newblock \emph{Journal of the Royal Statistical Society: Series B (Statistical
  Methodology)} \textbf{74}, 419 (2012).

\bibitem[{Brown and Sethna(2003)}]{Brown2003:StatisticalMechanicalApproaches}
K.~S. Brown and J.~P. Sethna.
\newblock Statistical mechanical approaches to models with many poorly known
  parameters.
\newblock \emph{Physical Review E} \textbf{68}, 021904 (2003).

\bibitem[{Gutenkunst et~al.(2007)Gutenkunst, Waterfall, Casey
  et~al.}]{Gutenkunst2007:UniversallySloppyParameter}
R.~N. Gutenkunst, J.~J. Waterfall, F.~P. Casey, et~al.
\newblock Universally sloppy parameter sensitivities in systems biology models.
\newblock \emph{PLoS Computational Biology} \textbf{3}, e189 (2007).

\bibitem[{Fulcher and Lubba(2020)}]{Zenodo_TimeSeriesGeneration}
B.~Fulcher and C.~H. Lubba.
\newblock benfulcher/timeseriesgeneration: v0.1 (2020).

\bibitem[{Fox(1987)}]{Fox1987}
C.~G. Fox.
\newblock {An inverse fourier transform algorithm for generating random signals
  of a specified spectral form}.
\newblock \emph{Computers and Geosciences} \textbf{13}, 369 (1987).

\end{thebibliography}
\end{document}